\pdfoutput=1
\documentclass{IEEEtran}
\usepackage{amsfonts}
\usepackage{graphicx}
\usepackage{textcomp}
\usepackage{xcolor}
\usepackage{verbatim}
\usepackage{gensymb}
\usepackage{srcltx}
\usepackage{psfrag}
\usepackage{epsfig}
\usepackage{color}
\usepackage[tbtags,sumlimits,nointlimits,reqno]{amsmath}
\usepackage{amsmath,amssymb,amsfonts}
\usepackage{color}
\usepackage{float}
\usepackage{mathtools, cuted}
\usepackage[makeroom]{cancel}
\usepackage[normalem]{ulem}
\usepackage{url}
\usepackage{cite}
\usepackage{multirow}

\definecolor{burntorange}{rgb}{0.8, 0.28, 0.0}
\definecolor{myGreen}{rgb}{0.0, 0.5, 0.0}
\definecolor{amber}{rgb}{0.8, 0.28, 0.0}
\definecolor{ceruleanblue}{rgb}{0.16, 0.28, 0.75}
\definecolor{matlabblue}{rgb}{0, 0.4470, 0.7410}
\definecolor{matlaborange}{rgb}{0.8500, 0.3250, 0.0980}
\definecolor{mypurple}{rgb}{0.314, 0.271, 0.588}

\usepackage{enumitem}
\usepackage{overpic}
\usepackage[crop=pdfcrop]{pstool}
\usepackage{color}
\usepackage{mathdots}
\usepackage{color}
\usepackage{mathtools, cuted}
\usepackage{bbm}
\usepackage{psfrag}
\usepackage{dblfloatfix}    
\usepackage{bbm}
\usepackage{array}
\usepackage{float}
\usepackage{srcltx}
\usepackage{color}
\usepackage{mathdots}
\usepackage[export]{adjustbox}
\usepackage[normalem]{ulem} 
\usepackage{color}
\usepackage{float}
\usepackage{mathtools, cuted}
\usepackage[makeroom]{cancel}
\usepackage{relsize}
\usepackage{bbm}
\usepackage{ulem}
\usepackage{flushend}

\makeatletter
\let\MYcaption\@makecaption
\makeatother

\usepackage[font=footnotesize]{subcaption}

\makeatletter
\let\@makecaption\MYcaption
\makeatother

\begin{document}
	
	\title{Integrated Multi-Port Leaky-Wave Antenna Multiplexer/Demultiplexer System for Millimeter-Wave Communication}
	\author{Mohamed K. Emara, \IEEEmembership{Graduate Student Member, IEEE} and Shulabh Gupta, \IEEEmembership{Senior Member, IEEE}
	\thanks{M. K. Emara and S. Gupta are with the Department of Electronics at Carleton University, Ottawa, ON K1S 5B6 Canada (email: mohamed.emara@carleton.ca)}}
	
	\maketitle
	
\begin{abstract}
	
	A novel application of leaky-wave antennas (LWAs) as integrated antennas and multiplexers/demultiplexers is proposed and experimentally demonstrated in the millimeter-wave band at 60~GHz. The first application is demultiplexing of an oblique-incident free-space wideband plane wave into $2N$ channels using $N$ LWAs with different beam-scanning laws. The second application is $N$--channel multiplexing and demultiplexing using $N$ pairs of identical LWAs in each other's far-field, where each LWA pair is designed such that the broadside frequency is the center frequency of its respective channel. Friis transmission equation is used to analytically demonstrate the two proposed applications using Gaussian radiation beams. The LWAs are then implemented using reflection-cancelling slot pairs using substrate integrated waveguide (SIW) technology. Full-wave simulations are used to demonstrate the first application by demultiplexing an incident plane wave into two and four channels using two LWAs. Finally, the two proposed applications are experimentally demonstrated by conducting horn-to-LWA and LWA-to-LWA transmission measurements and successfully showing broadband frequency discrimination. The proposed LWAs provide a simple, compact, high-efficiency, and low-profile multiplexing/demultiplexing solution with high-integration capability with other circuitry. Compared to conventional multiplexers, the proposed solution does not require matching networks and can be directly scaled to higher frequencies using the same architecture.
	
\end{abstract}



\section{Introduction}
	Multiplexers and demultiplexers are fundamental components in wireless communication systems. They are used to separate frequencies in a wideband signal from an antenna into distinct bands (or channels) in the front-end \cite{Cameron_MUX}. For instance, they are used in multiband wireless applications such as satellite communications \cite{Kunes_MUX, Yun_MUX, Hu_MUX, Chang_MUX}, global navigation satellite systems (GNSS) receivers \cite{Deng_MUX, Oshima_MUX_SAW, Vladimirov_MUX}, meteorological satellites \cite{Zhang_MUX}, and other wireless applications \cite{Lai_MUX, Choi_MUX, Lee_MUX, Weng_MUX}. The upcoming fifth generation (5G) communication networks are naturally expected to see a high demand for high-efficiency multiplexing and demultiplexing solutions in the millimeter-wave (mm-wave) band. For instance, the IEEE 802.11ad standard at $60$~GHz consists of six channels between $57.24$ and $70.20$~GHz. Each channel occupies $2.16$~GHz (including guard bands) and provides $1.76$~GHz of bandwidth \cite{Wu_IEEE802} \cite{Nitsche_IEEE802}. It is therefore expected that there will be a need for novel solutions that can provide multiplexing and demultiplexing of channels at these frequencies.

	Multiplexers are conventionally implemented using bandpass filters (BPF) combined with matching networks to achieve low reflection at the common port. For example, multiplexers have been designed using microstrip bandpass filters \cite{Weng_MUX, Yun_MUX, Matthaei_MUX}. Microstrip-based multiplexers are planar, compact and easy to integrate with other circuit elements on a printed circuit board (PCB), however they suffer from high losses and low quality factors. Multiplexers have also been implemented using waveguide T-junctions and waveguide filters \cite{Yao_MUX, Mohottige_MUX}. Waveguide-based multiplexers exhibit low loss, high quality factors, and high power handling capabilities, however they do not integrate well with other circuitry and are usually bulky. To overcome these challenges, substrate-integrated waveguide (SIW) multiplexers have been designed \cite{Hao_MUX, Dong_MUX} which offer the advantages of waveguide techniques while maintaining a planar design that is easy to integrate with other circuitry.
	
	Other implementations of multiplexers include using surface acoustic wave (SAW) filters \cite{Solie_MUX} and lumped-element bandpass filters \cite{Ahn_MUX} on PCBs. While these techniques are simple to implement from a fabrication point-of-view, they typically exhibit high losses and challenging matching at the common port. In addition, SAW filters have a limited usage up to about $3$~GHz, and lumped-element filters are challenging to scale to high frequencies, so neither are suitable for implementation of mm-wave multiplexers. In the mm-wave band, a multiplexer was designed for channels around $100$~GHz using a blazed diffraction grating \cite{Henry_MUX}, but it involves a complex apparatus and fabrication using machine-milled aluminum plates. Another multiplexing solution in the $70$ to $80$~GHz band based on parallel-plate Fabry-P\'erot resonators was proposed in \cite{Arnaud_MUX}, however the design is bulky and does not integrate well with PCB fabrication and circuitry. As opposed to an integrated solution, a free-space diplexer solution around $2.5$~GHz based on composite right/left-handed (CRLH) leaky-wave antenna was proposed in \cite{Gupta_MUX}. Around the same time, a duplexer/diplexer CRLH ferrite-loaded open waveguide was proposed in \cite{Kodera_MUX}. Both solutions utilized the frequency-scanning property of leaky-wave antennas. They were however limited to only two channels, and demonstrated within the microwave frequency bands,

\begin{figure*}[!h]
	\centering
	\begin{subfigure}{1\textwidth}
		\centering
		\begin{overpic}[grid=false, scale = 0.48]{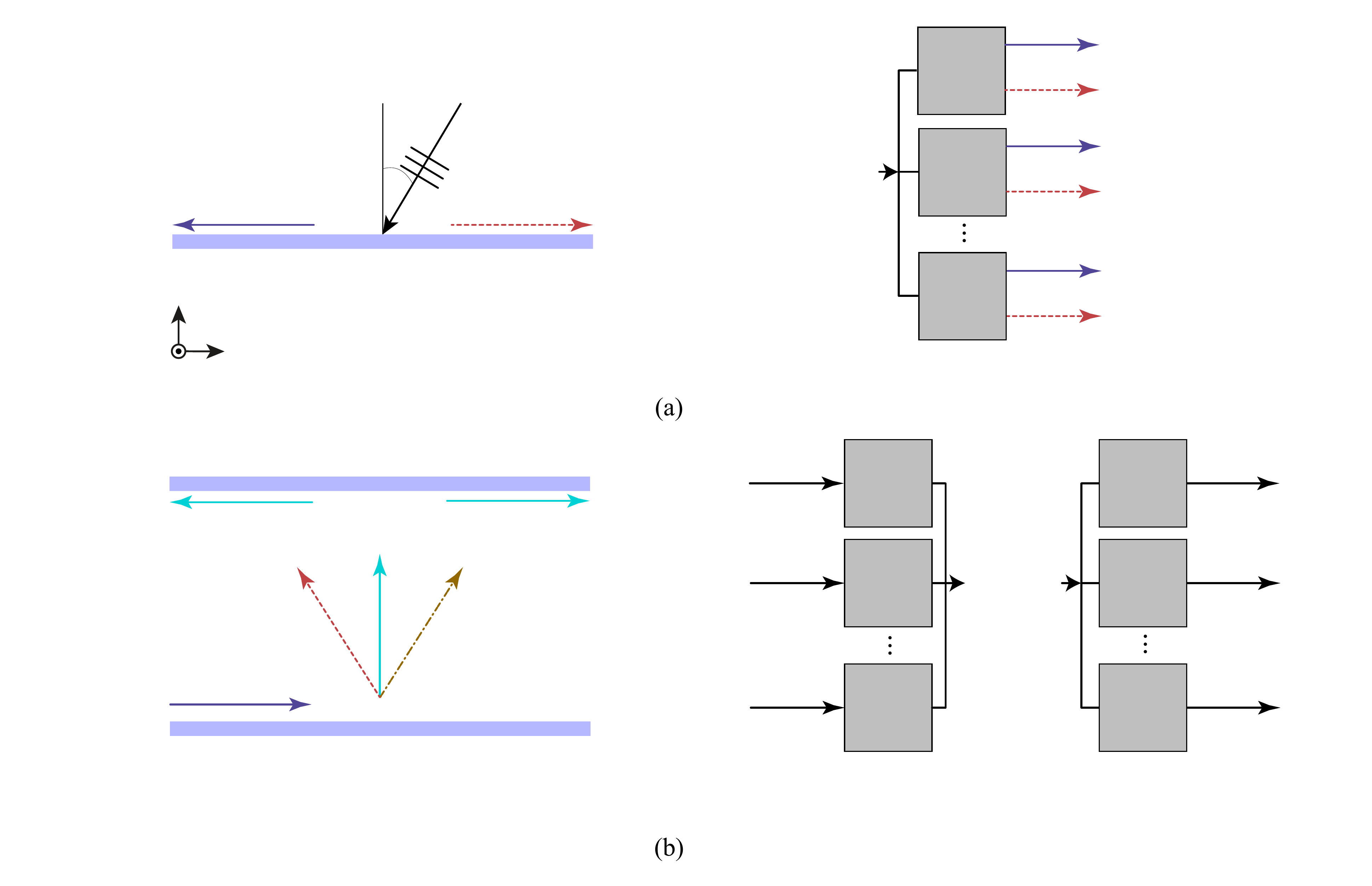}
			\put(10,47){\makebox(0,0){\scriptsize Port 1}}
			\put(10,45.5){\makebox(0,0){\scriptsize (Rx)}}
			
			\put(46,47){\makebox(0,0){\scriptsize Port 2}}
			\put(46,45.5){\makebox(0,0){\scriptsize (Rx)}}
			
			\put(10,29.5){\makebox(0,0){\scriptsize Port 1}}
			\put(10,28){\makebox(0,0){\scriptsize (Rx)}}
			
			\put(46,29.5){\makebox(0,0){\scriptsize Port 2}}
			\put(46,28){\makebox(0,0){\scriptsize (matched)}}
			
			\put(10,11.5){\makebox(0,0){\scriptsize Port 1}}
			\put(10,10){\makebox(0,0){\scriptsize (Tx)}}
			
			\put(46,11.5){\makebox(0,0){\scriptsize Port 2}}
			\put(46,10){\makebox(0,0){\scriptsize (matched)}}
			
			\put(30.2,55){\makebox(0,0){\scriptsize $\theta\ne0\degree$}}
			\put(34,58){\makebox(0,0){\scriptsize $k$}}
			
			\put(13,37.5){\makebox(0,0){\scriptsize $x$}}		
			\put(17.2,39){\makebox(0,0){\scriptsize $y$}}			
			\put(13,43){\makebox(0,0){\scriptsize $z$}}
			
			\put(34,60){\makebox(0,0){\scriptsize\color{blue}\shortstack{\textsc{Input}\\\textsc{Plane Wave}}}}
			
			\put(38,49){\makebox(0,0){\scriptsize\color{purple}\shortstack{\textsc{Backward}}}}
			\put(38,50.5){\makebox(0,0){\scriptsize $C_\text{out}(f_\text{c} = f_1)$}}
			
			\put(18,49){\makebox(0,0){\scriptsize\color{mypurple}\shortstack{\textsc{Forward}}}}
			\put(18,50.5){\makebox(0,0){\scriptsize $C_\text{out}(f_\text{c} = f_1)$}}
			
			\put(76.5,62){\makebox(0,0){\scriptsize Port 1}}
			\put(85,61.3){\makebox(0,0){\scriptsize $C_\text{out}(f_\text{c} = f_1)$}}
			\put(76.5,57.5){\makebox(0,0){\scriptsize Port 2}}
			\put(85,58){\makebox(0,0){\scriptsize $C_\text{out}(f_\text{c} = f_2)$}}	
			
			\put(76.5,54.5){\makebox(0,0){\scriptsize Port 3}}
			\put(85,53.8){\makebox(0,0){\scriptsize $C_\text{out}(f_\text{c} = f_3)$}}
			\put(76.5,50){\makebox(0,0){\scriptsize Port 4}}
			\put(85,50.5){\makebox(0,0){\scriptsize $C_\text{out}(f_\text{c} = f_4)$}}	
			
			\put(77,45.8){\makebox(0,0){\scriptsize Port 2$N$-1}}
			\put(85.5,44.8){\makebox(0,0){\scriptsize $C_\text{out}(f_\text{c} = f_{2N\text{-}1})$}}
			\put(77,40.5){\makebox(0,0){\scriptsize Port 2$N$}}
			\put(85,41.5){\makebox(0,0){\scriptsize $C_\text{out}(f_\text{c} = f_{2N})$}}	
			
			\put(18,14.5){\makebox(0,0){\scriptsize\color{mypurple}\shortstack{\textsc{Input Channel}}}}
			\put(18,16){\makebox(0,0){\scriptsize $C_\text{in}(f_\text{c} = f_\text{bs})$}}
			\put(18,27){\makebox(0,0){\scriptsize $C_\text{out}(f_\text{c} = f_\text{bs})$}}
			\put(27.5,25){\makebox(0,0){\scriptsize $f_\text{bs}$}}		
			\put(21,24){\makebox(0,0){\scriptsize $f_\text{b}$}}		
			\put(35,24){\makebox(0,0){\scriptsize $f_\text{f}$}}		
			
			\put(70,38){\makebox(0,0){\scriptsize\color{myGreen}\shortstack{\textsc{Demultiplexer}}}}
			\put(83,8){\makebox(0,0){\scriptsize\color{myGreen}\shortstack{\textsc{Demultiplexer}}}}
			\put(65,8){\makebox(0,0){\scriptsize\color{myGreen}\shortstack{\textsc{Multiplexer}}}}
			
			\put(73.8,22.1){\makebox(0,0){\scriptsize\color{black}\shortstack{\textsc{Free-Space}\\\textsc{Plane-Wave}}}}
			\put(60.5,52.1){\makebox(0,0){\scriptsize\color{black}\shortstack{\textsc{Free-Space}\\\textsc{Plane-Wave}}}}
			
			\put(27.5,45.5){\makebox(0,0){\scriptsize LWA}}
			\put(27.5,31){\makebox(0,0){\scriptsize LWA}}
			\put(27.5,10){\makebox(0,0){\scriptsize LWA}}
			
			\put(70.3,59.5){\makebox(0,0){\scriptsize LWA 1}}
			\put(70.3,52){\makebox(0,0){\scriptsize LWA 2}}
			\put(70.3,43){\makebox(0,0){\scriptsize LWA $N$}}
			
			\put(83.5,29.3){\makebox(0,0){\scriptsize LWA 1}}
			\put(83.5,22.1){\makebox(0,0){\scriptsize LWA 2}}
			\put(83.5,13.1){\makebox(0,0){\scriptsize LWA $N$}}
			
			\put(64.7,29.3){\makebox(0,0){\scriptsize LWA 1}}
			\put(64.7,22.1){\makebox(0,0){\scriptsize LWA 2}}
			\put(64.7,13.1){\makebox(0,0){\scriptsize LWA $N$}}
			
			\put(4,48){\makebox(0,0){\rotatebox{90}{\scriptsize\color{burntorange}\shortstack{\textsc{Plane Wave 2N-Channel}\\\textsc{Demultiplexer Using N-LWAs}}}}}
			\put(4,20){\makebox(0,0){\rotatebox{90}{\scriptsize\color{burntorange}\shortstack{\textsc{N-Channel Multiplexer/}\\\textsc{Demultiplexer Using N-LWAs}}}}}
			
			\put(57,31){\makebox(0,0){\scriptsize Port 1 (Tx)}}
			\put(57,28){\makebox(0,0){\scriptsize $C_\text{in}(f_\text{c} = f_\text{bs1})$}}
			
			\put(57,23.5){\makebox(0,0){\scriptsize Port 3 (Tx)}}
			\put(57,20.5){\makebox(0,0){\scriptsize $C_\text{in}(f_\text{c} = f_\text{bs2})$}}
			
			\put(57,14){\makebox(0,0){\scriptsize Port 2$N$-1 (Tx)}}
			\put(57,11){\makebox(0,0){\scriptsize $C_\text{in}(f_\text{c} = f_{\text{bs}N})$}}
			
			\put(92,31){\makebox(0,0){\scriptsize Port 1 (Rx)}}
			\put(92,28){\makebox(0,0){\scriptsize $C_\text{out}(f_\text{c} = f_\text{bs1})$}}
			
			\put(92,23.5){\makebox(0,0){\scriptsize Port 3 (Rx)}}
			\put(92,20.5){\makebox(0,0){\scriptsize $C_\text{out}(f_\text{c} = f_\text{bs2})$}}
			
			\put(92,14){\makebox(0,0){\scriptsize Port 2$N$-1 (Rx)}}
			\put(92,11){\makebox(0,0){\scriptsize $C_\text{out}(f_\text{c} = f_{\text{bs}N})$}}
		\end{overpic}
	\end{subfigure}%
	
	\caption{Illustrations of the proposed multiplexing and demultiplexing applications of $N$ LWAs with different beam-scanning laws. a) Demultiplexing a free-space broadband plane wave, incident at an angle $\theta$, into $2N$ channels using $N$ LWAs. b) Multiplexing of $N$ channels input at $N$ LWAs radiating at the broadside frequencies (every broadside frequency is the center frequency of the respective channel), and demultiplexing of the same channels at identical $N$ LWAs in the far-field from free-space plane waves to recover the original $N$ channels.}
	\label{fig:LWA_Proposed}
\end{figure*}
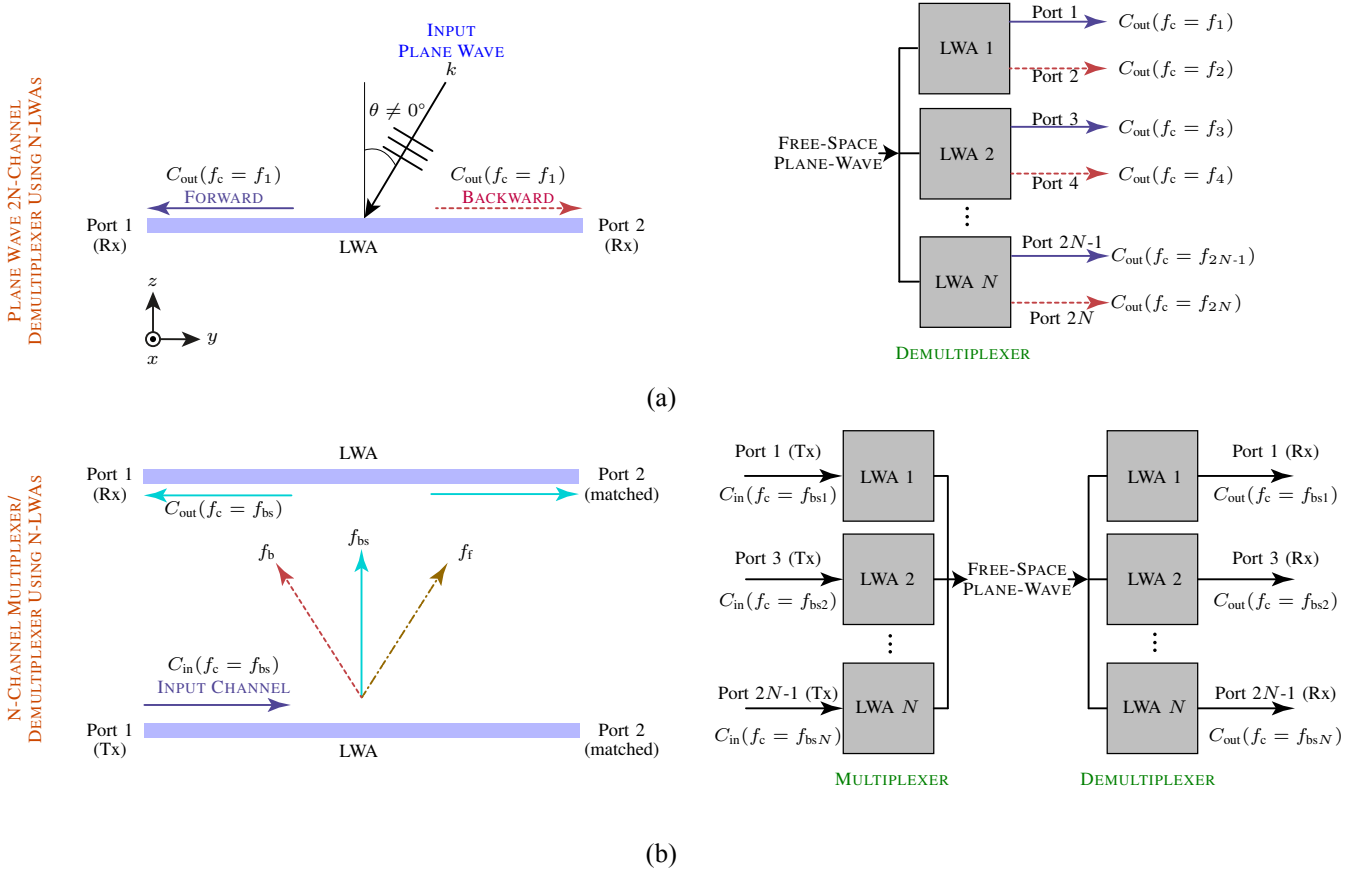

	 In this paper, we propose and demonstrate a novel free-space multiplexer and demultiplexer technique using a multi-port antenna system consisting of an array of engineered leaky wave antennas (LWAs). Exploiting their frequency-dependent scanning property, multiple channel frequencies are automatically combined into free-space and radiated along a desired direction, i.e. channel multiplexer. Conversely, the multi-port antenna can receive multiplexed channels along a given direction and route individual channels to unique antenna ports. Therefore, it acts as a frequency discriminator and performs an analog channel demultiplexing based on frequency. The proposed solution is based on substrate-integrated waveguides (SIWs) and implemented using printed circuit board (PCB) technology, so it is compact, planar, and low-cost. The proposed design furthermore does not need matching network and so it is simpler to implement than conventional multiplexers and can be directly scaled to higher frequencies using the same architecture. Finally, the proposed design can be easily integrated with other circuit elements, due to its SIW implementation, such as low-noise amplifiers (LNAs) in the communication system front-end. The basic idea was proposed in \cite{Emara_LWA} and in this paper the concept is further expanded upon by analytical demonstration using Friis transmission equation and experimental demonstration in the mm-wave band at $60$~GHz using SIW slot array antennas.
	
	The paper is organized as follows. Sec.~II presents the proposed multiplexer/demultiplexer principle and provides an analytical demonstration using Friis transmission equation. Sec.~III provides a full-wave demonstration of demultiplexing of normal- and oblique-incidence plane waves into discrete channels using slot array LWAs. Next, Sec.~IV presents an experimental demonstration of demultiplexing using horn-to-LWA transmission, and multiplexing/demultiplexing using LWA-to-LWA transmission. Sec.~V provides a general discussion and an investigation of the bandwidth of the proposed design. Conclusions are finally provided in Sec.~VI.

\section{Proposed LWA Multiplexer/Demultiplexer}
\subsection{Proposed Principle}
	
	Leaky-wave antennas (LWAs) are typically two-port traveling-wave structures in which the electromagnetic power radiates as it travels along the guiding structure. Two applications of LWAs are proposed as shown in Fig.~\ref{fig:LWA_Proposed}. The first application is demultiplexing of an oblique-incident wideband plane wave signal into $2N$ channels using $N$ LWAs as shown in Fig.~\ref{fig:LWA_Proposed}(a). The second application is multiplexing and demultiplexing $N$ channels using $N$ LWAs as shown in Fig.~\ref{fig:LWA_Proposed}(b). 
	 
	These operations are possible due to the frequency-selectivity of LWAs. Frequency-selective antennas are needed in order to perform multiplexing/demultiplexing at the antenna level without the need for separate multiplexers and demultiplexers. While most antennas such as dipoles, patches, and horns radiate all frequencies in a certain direction \cite{Balanis}, a LWA's main beam angle scans with frequency from backward to forward region including broadside radiation \cite{Collin_antenna, Jackson_LWA,Otto_LWA, Xu_LWA,Hines_Slot}.  A LWA can be characterized by its frequency-dependent complex propagation constant $\gamma(\omega)=\alpha(\omega) + j\beta(\omega)$, where $\alpha$ is the leakage per unit length, and $\beta$ is the propagation constant per unit length, along the antenna structure. LWA are electrically large and feature high efficiency and high gain without the need for complex feeding networks. The phase shift between the LWA unit cells is accounted for by $\beta(\omega)$ and it is used to determine the main beam direction, $\theta(\omega)$ measured from the broadside axis (e.g. $z-$axis), in the beam-scanning law
	 
	\begin{equation}
	\theta(\omega) = \sin^{-1}\left[\frac{\beta(\omega)}{k_0}\right],
	\label{Eq:Law}
	\end{equation} 
	
	\noindent where $k_0=\omega/c$ is the free-space wavenumber. 
	
	Since, a LWA is a 2-port antenna, it can be excited from either of the ports exhibiting the beam-scanning law with respect to the excitation port, i.e. with increasing frequency, the main beam of a LWA scans from the backward region through broadside to the forward region, \textit{with respect to the feeding port}. For instance, consider a forward radiating frequency $f_1$ along $\theta$, when Port 1 of the LWA is excited. Simultaneously, a different frequency $f_2$ may also be chosen which, when excited from Port 2, can also radiate along the same angle $\theta$. Therefore, if two channel frequencies $\{f_1,~f_2\}$ exciting LWA Ports $\{1,~2\}$, respectively, radiate along the same direction $\theta$, this effectively \emph{multiplexes} the two channels, i.e. $2:1$ multiplexer. Since the antenna is reciprocal, these two channel frequencies incident on a 2-port LWA apertures instead, will be discriminated and individually routed to the two ports, as illustrated in Fig.~\ref{fig:LWA_Proposed}(a), i.e. $1:2$ demultiplexer. This idea can now be extended to more than 2 channels, where by using $N$~LWAs with different beam-scanning laws, the resulting $2N$~ports may be engineered to discriminate $2N$ channel frequencies illuminating the multi-port antenna aperture from a fixed \textit{oblique} direction in space, as shown in Fig.~\ref{fig:LWA_Proposed}(a). This forms the proposed \textit{multiplexing/demultiplexing solution} using multi-port LWAs as either transmitting or receiving antennas. If the LWA aperture is designed to be long, the two ports are fully isolated. An exception is a broadside frequency corresponding to $\theta=0$, in whereby an incoming broadside frequency is equally divided between the two ports. In this case, the two ports are not isolated anymore, and the LWA fails to operate as a multiplexer or a demultiplexer.  
	
	Naturally, in the above case, the receiving frequency pairs assigned at each antenna are related by the beam scanning law of the antenna, and the antenna may not receive from broadside direction. These conditions may be relaxed if two \emph{identical LWAs} of finite sizes are used as a Tx-Rx pair operating at broadside frequencies under line-of-sight (LOS) far-field radiation conditions, as shown in Fig.~\ref{fig:LWA_Proposed}(b). Consider a three-tone signal with frequencies corresponding to backward ($f_\text{b}$), broadside ($f_\text{bs}$) and forward ($f_\text{f}$) radiation regions, exciting Port 1 of the Tx antenna. The receive (Rx) LWA will receive peak power at the broadside frequency (at Port 1 and Port 2, equally), rolling off at the forward and backward frequencies based on the beam-scanning law and the beamwidth of the antenna. If $f_\text{b}$ and $f_\text{f}$ are chosen such that the power received by the Rx LWA is $-3$~dB compared to the peak radiation at broadside $f_\text{bs}$, $\Delta f = |f_\text{f} - f_\text{b}|$ may be considered as a channel bandwidth that this Tx-Rx LWA pair can support. Therefore, for a given channel bandwidth $\Delta f$ centered around $\approx f_\text{bs}$, these two LWAs form a unique Tx-Rx pair. This idea can now be extended to $N$ adjacent channels using $N$ LWA pairs (one Tx and one Rx for each channel), where each pair has a unique beam-scanning law, with different broadside frequencies. As a result, each Rx LWA is only sensitive to the channel centered at its broadside frequency, while being isolated to all other channels. The resulting array of $N$ antennas forming the multi-port antenna structure with $2N$ ports thus acts as a channel multiplexer when used a transmitter, and a channel demultiplexer when used as a receiver. This finally forms the proposed \textit{multiplexing/demultiplexing solution} using LWA pairs in a Tx-Rx configuration.

\subsection{Analytical Demonstration Using Friis Equation}

\begin{figure*}[!t]
	\centering
	\begin{subfigure}{1\textwidth}
		\centering
		\begin{overpic}[grid=false, scale = 0.35]{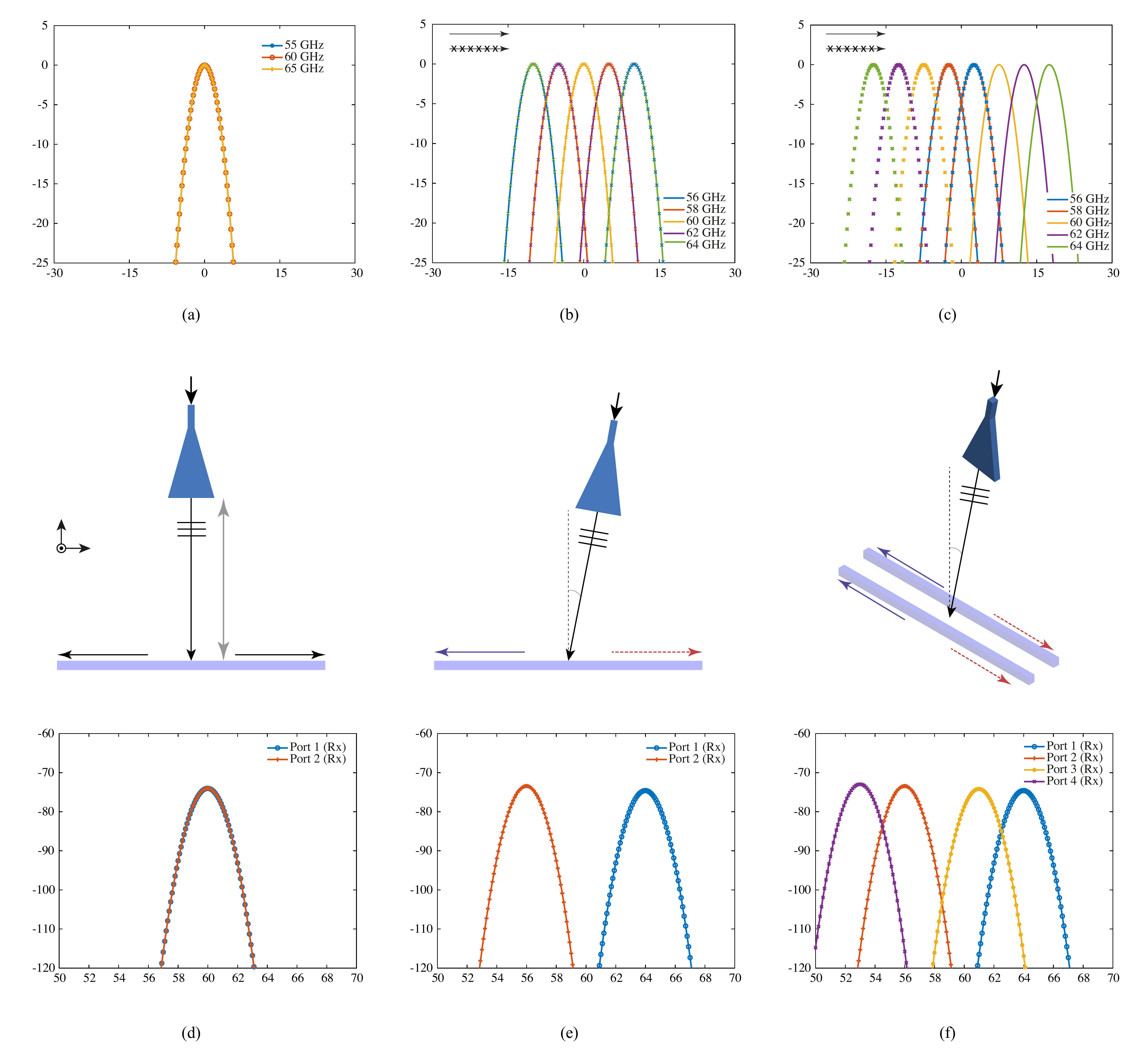}
			\put(17.8,67.5){\makebox(0,0){\scriptsize Theta, $\theta(\degree)$}}
			\put(2,80){\makebox(0,0){\scriptsize\rotatebox{90}{Gain (dBi)}}}
			
			\put(51,67.5){\makebox(0,0){\scriptsize Theta, $\theta(\degree)$}}
			\put(35,80){\makebox(0,0){\scriptsize\rotatebox{90}{Gain (dBi)}}}
			
			\put(85,67.5){\makebox(0,0){\scriptsize Theta, $\theta(\degree)$}}
			\put(68,80){\makebox(0,0){\scriptsize\rotatebox{90}{Gain (dBi)}}}
			
			\put(45,89){{\scriptsize Port 1 (Increasing Frequency)}}
			\put(45,87.7){{\scriptsize Port 2 (Decreasing Frequency)}}
			
			\put(78,89){{\scriptsize Port 1 (Increasing Frequency)}}
			\put(78,87.7){{\scriptsize Port 2 (Decreasing Frequency)}}
			
			\put(16.7,61){{\makebox(0,0){\scriptsize $P_\text{t}$}}}
			\put(54.2,59){{\makebox(0,0){\scriptsize $P_\text{t}$}}}
			\put(87,61){{\makebox(0,0){\scriptsize $P_\text{t}$}}}
			
			\put(23,43){\makebox(0,0){\scriptsize $R = 2$~m}}		
			\put(5.3,43.5){\makebox(0,0){\scriptsize $x$}}		
			\put(8.5,44.5){\makebox(0,0){\scriptsize $y$}}			
			\put(5.3,48){\makebox(0,0){\scriptsize $z$}}
			
			\put(16.7,33){\makebox(0,0){\scriptsize LWA 1}}
			\put(10,36.5){\makebox(0,0){\scriptsize\color{black}\shortstack{\textsc{Broadside}}}}
			\put(24.5,36.5){\makebox(0,0){\scriptsize\color{black}\shortstack{\textsc{Broadside}}}}
			
			\put(3,35){\makebox(0,0){\scriptsize Port 1}}
			\put(3,33.5){\makebox(0,0){\scriptsize (Rx)}}
			
			\put(31,35){\makebox(0,0){\scriptsize Port 2}}
			\put(31,33.5){\makebox(0,0){\scriptsize (Rx)}}
			
			\put(50,33){\makebox(0,0){\scriptsize LWA 1}}
			\put(42,36.5){\makebox(0,0){\scriptsize\color{mypurple}\shortstack{\textsc{Forward}}}}
			\put(57,36.5){\makebox(0,0){\scriptsize\color{purple}\shortstack{\textsc{Backward}}}}
			\put(53.5,40){\makebox(0,0){\scriptsize $\theta=10\degree$}}
			
			\put(35.5,35){\makebox(0,0){\scriptsize Port 1}}
			\put(35.5,33.5){\makebox(0,0){\scriptsize (Rx)}}
			
			\put(63.5,35){\makebox(0,0){\scriptsize Port 2}}
			\put(63.5,33.5){\makebox(0,0){\scriptsize (Rx)}}
			
			\put(87,47){\makebox(0,0){\scriptsize $\theta=10\degree$}}
			\put(79.5,36.5){\makebox(0,0){\scriptsize LWA 1}}
			\put(86,41){\makebox(0,0){\scriptsize LWA 2}}
			\put(76,39){\makebox(0,0){\rotatebox{-30}{\scriptsize\color{mypurple}\shortstack{\textsc{Forward}}}}}
			\put(80,44){\makebox(0,0){\rotatebox{-30}{\scriptsize\color{mypurple}\shortstack{\textsc{Forward}}}}}
			\put(84.5,34){\makebox(0,0){\rotatebox{-30}{\scriptsize\color{purple}\shortstack{\textsc{Backward}}}}}
			\put(90,38){\makebox(0,0){\rotatebox{-30}{\scriptsize\color{purple}\shortstack{\textsc{Backward}}}}}

			\put(71,44.5){\makebox(0,0){\scriptsize Port 1}}
			\put(71,43){\makebox(0,0){\scriptsize (Rx)}}
			
			\put(75,47){\makebox(0,0){\scriptsize Port 3}}
			\put(75,45.5){\makebox(0,0){\scriptsize (Rx)}}
			
			\put(90,31.5){\makebox(0,0){\scriptsize Port 2}}
			\put(90,30){\makebox(0,0){\scriptsize (Rx)}}
			
			\put(94,33.5){\makebox(0,0){\scriptsize Port 4}}
			\put(94,32){\makebox(0,0){\scriptsize (Rx)}}
			
			\put(17.8,6){\makebox(0,0){\scriptsize Frequency (GHz)}}
			\put(2,18){\makebox(0,0){\scriptsize\rotatebox{90}{Power received (dBm)}}}
			
			\put(51,6){\makebox(0,0){\scriptsize Frequency (GHz)}}
			\put(35,18){\makebox(0,0){\scriptsize\rotatebox{90}{Power received (dBm)}}}
			
			\put(85,6){\makebox(0,0){\scriptsize Frequency (GHz)}}
			\put(68,18){\makebox(0,0){\scriptsize\rotatebox{90}{Power received (dBm)}}}
			
		\end{overpic}
	\end{subfigure}
	
	\caption{Horn-to-LWA analytical demonstration using Friis transmission equation. Radiations patterns of the antennas used in this demonstration : a) radiation patterns of the horn at $59$, $60$ and $61$~GHz with the horn pointing in the broadside direction (for an angle of $10\degree$ from broadside, the peak simply shifts to $\theta=10\degree$), b) radiation patterns of LWA \#1 from $58$ to $62$~GHz for excitation from Port 1 and Port 2, and c) radiation patterns of LWA \#2 from $58$ to $62$~GHz for excitation from Port 3 and Port 4. The Friis transmission calculation was done for three cases: d) horn at broadside with respect to LWA \#1, e) horn at $\theta=10\degree$ with respect to LWA \#1, and f) horn at $\theta=10\degree$ with respect to LWA \#1 and LWA \#2. In all cases, the transmitted power is $0$~dBm across all frequencies. }
	\label{fig:horn_LWA_Friis}
\end{figure*}

\begin{figure*}[!b]
	\centering
		\begin{subfigure}{1\textwidth}
		\centering
		\begin{overpic}[grid=false, scale = 0.31]{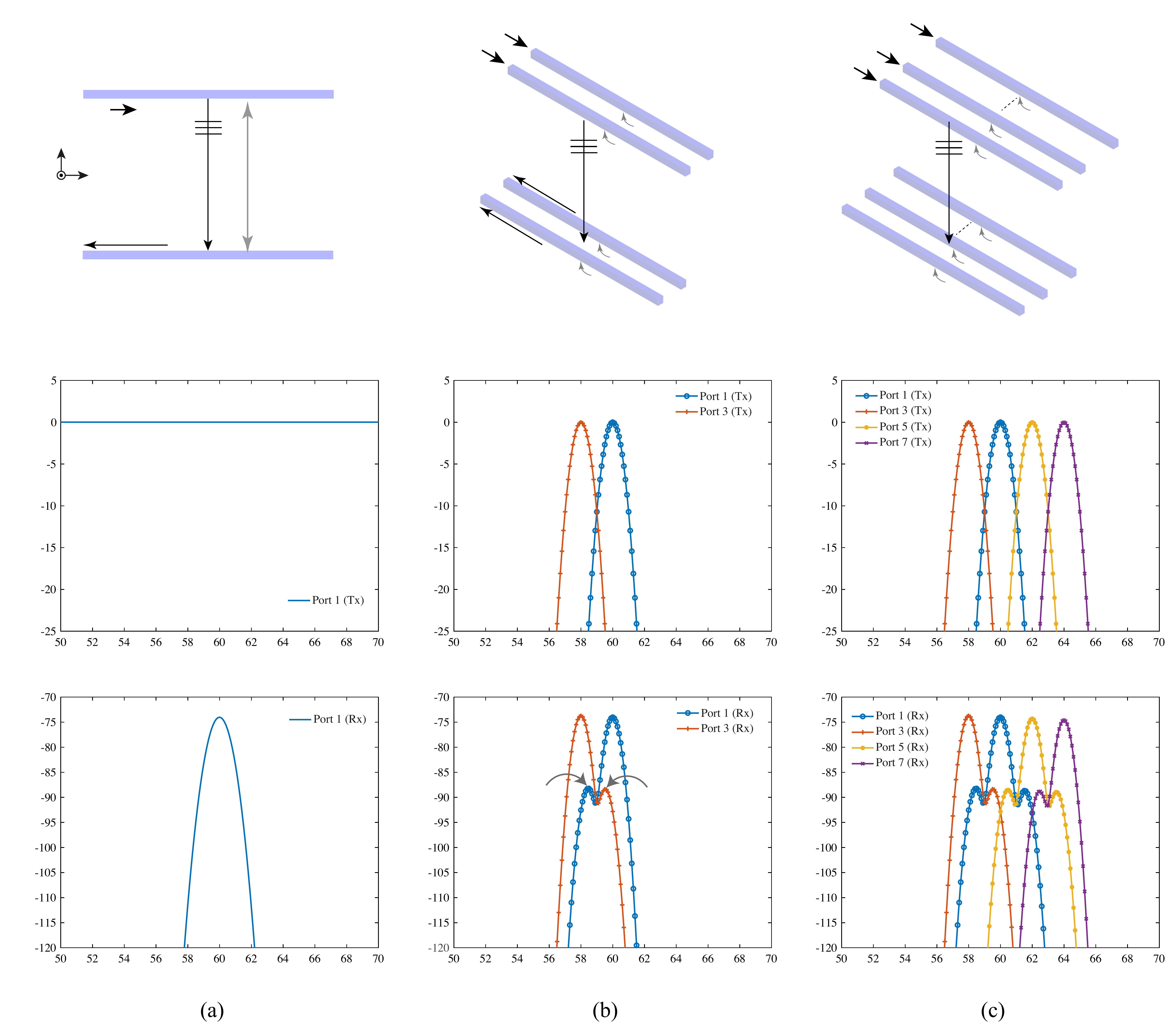}
			\put(17.5,65){\makebox(0,0){\scriptsize\color{burntorange} LWA 1}}
			\put(17.5,81){\makebox(0,0){\scriptsize\color{burntorange} LWA 1}}
			
			\put(8,78.5){\makebox(0,0){\scriptsize $P_\text{t}$}}
			
			\put(11,77.5){\makebox(0,0){\scriptsize\color{black}\shortstack{\textsc{Input}}}}
			\put(11,68.5){\makebox(0,0){\scriptsize\color{black}\shortstack{\textsc{Broadside}}}}
						
			\put(4.5,80.5){\makebox(0,0){\scriptsize Port 1}}
			\put(4.5,79){\makebox(0,0){\scriptsize (Tx)}}
			
			\put(31.4,80.5){\makebox(0,0){\scriptsize Port 2}}
			\put(31.4,79){\makebox(0,0){\scriptsize (matched)}}
			
			\put(4.5,66.5){\makebox(0,0){\scriptsize Port 1}}
			\put(4.5,65){\makebox(0,0){\scriptsize (Rx)}}
			
			\put(31.4,66.5){\makebox(0,0){\scriptsize Port 2}}
			\put(31.4,65){\makebox(0,0){\scriptsize (matched)}}
			
			\put(24.5,73){\makebox(0,0){\scriptsize $R = 2$~m}}		
			\put(5.2,71.5){\makebox(0,0){\scriptsize $x$}}		
			\put(8,73){\makebox(0,0){\scriptsize $y$}}			
			\put(5.2,76){\makebox(0,0){\scriptsize $z$}}
			
			\put(54,74.5){\makebox(0,0){\color{burntorange}\rotatebox{-30}{\scriptsize LWA 1}}}
			\put(55.7,76.2){\makebox(0,0){\color{burntorange}\rotatebox{-30}{\scriptsize LWA 2}}}
			
			\put(52,63.3){\makebox(0,0){\color{burntorange}\rotatebox{-30}{\scriptsize LWA 1}}}
			\put(53.7,65){\makebox(0,0){\color{burntorange}\rotatebox{-30}{\scriptsize LWA 2}}}
			
			\put(40,84){\makebox(0,0){\scriptsize $P_\text{t}$}}
			\put(42.5,86){\makebox(0,0){\scriptsize $P_\text{t}$}}
			
			\put(43.5,67.5){\makebox(0,0){\rotatebox{-30}{\scriptsize\color{black}\shortstack{\textsc{Broadside}}}}}
			\put(47.5,72){\makebox(0,0){\rotatebox{-30}{\scriptsize\color{black}\shortstack{\textsc{Broadside}}}}}
			
			\put(41,81){\makebox(0,0){\scriptsize Port 1}}
			\put(41,79.5){\makebox(0,0){\scriptsize (Tx)}}
		
			\put(47.8,85){\makebox(0,0){\scriptsize Port 3}}
			\put(47.8,83.5){\makebox(0,0){\scriptsize (Tx)}}
			
			\put(58,72){\makebox(0,0){\scriptsize Port 2}}
			\put(58,70.5){\makebox(0,0){\scriptsize (matched)}}
			
			\put(64,76){\makebox(0,0){\scriptsize Port 4}}
			\put(64,74.5){\makebox(0,0){\scriptsize (matched)}}
			
			\put(39,72){\makebox(0,0){\scriptsize Port 1}}
			\put(39,70.5){\makebox(0,0){\scriptsize (Rx)}}
			
			\put(43,75.5){\makebox(0,0){\scriptsize Port 3}}
			\put(43,74){\makebox(0,0){\scriptsize (Rx)}}
			
			\put(55,60.5){\makebox(0,0){\scriptsize Port 2}}
			\put(55,59){\makebox(0,0){\scriptsize (matched)}}
			
			\put(60,62.5){\makebox(0,0){\scriptsize Port 4}}
			\put(60,61){\makebox(0,0){\scriptsize (matched)}}
			
			\put(72,83){\makebox(0,0){\scriptsize $P_\text{t}$}}
			\put(74,84.5){\makebox(0,0){\scriptsize $P_\text{t}$}}
			\put(77,86.8){\makebox(0,0){\scriptsize $P_\text{t}$}}
			
			\put(73,80){\makebox(0,0){\scriptsize Port 1}}
			\put(73,78.5){\makebox(0,0){\scriptsize (Tx)}}
			
			\put(78.5,82.5){\makebox(0,0){\scriptsize Port 3}}
			\put(78.5,81){\makebox(0,0){\scriptsize (Tx)}}
			
			\put(84,85){\makebox(0,0){\scriptsize Port 2$N$-1}}
			\put(84,83.5){\makebox(0,0){\scriptsize (Tx)}}
			
			\put(90,70){\makebox(0,0){\scriptsize Port 2}}
			\put(90,68.5){\makebox(0,0){\scriptsize (matched)}}
			
			\put(94,72.5){\makebox(0,0){\scriptsize Port 4}}
			\put(94,71){\makebox(0,0){\scriptsize (matched)}}
			
			\put(98,76){\makebox(0,0){\scriptsize Port 2$N$}}
			\put(98,74.5){\makebox(0,0){\scriptsize (matched)}}
			
			\put(70,70){\makebox(0,0){\scriptsize Port 1}}
			\put(70,68.5){\makebox(0,0){\scriptsize (Rx)}}
			
			\put(72.5,72.5){\makebox(0,0){\scriptsize Port 3}}
			\put(72.5,71){\makebox(0,0){\scriptsize (Rx)}}
			
			\put(75,75.5){\makebox(0,0){\scriptsize Port 2$N$-1}}
			\put(75,74){\makebox(0,0){\scriptsize (Rx)}}
			
			\put(85,60){\makebox(0,0){\scriptsize Port 2}}
			\put(85,58.5){\makebox(0,0){\scriptsize (matched)}}
			
			\put(91,61.5){\makebox(0,0){\scriptsize Port 4}}
			\put(91,60){\makebox(0,0){\scriptsize (matched)}}
			
			\put(95,65){\makebox(0,0){\scriptsize Port 2$N$}}
			\put(95,63.5){\makebox(0,0){\scriptsize (matched)}}
			
			\put(86,73){\makebox(0,0){\color{burntorange}\rotatebox{-30}{\scriptsize LWA 1}}}
			\put(87.5,74.8){\makebox(0,0){\color{burntorange}\rotatebox{-30}{\scriptsize LWA 2}}}
			\put(89.5,77){\makebox(0,0){\color{burntorange}\rotatebox{-30}{\scriptsize LWA $N$}}}
			
			\put(82.5,62.7){\makebox(0,0){\color{burntorange}\rotatebox{-30}{\scriptsize LWA 1}}}
			\put(84.2,64.3){\makebox(0,0){\color{burntorange}\rotatebox{-30}{\scriptsize LWA 2}}}
			\put(86.5,65.7){\makebox(0,0){\color{burntorange}\rotatebox{-30}{\scriptsize LWA $N$}}}
			
			\put(18.5,32){\makebox(0,0){\scriptsize Frequency (GHz)}}
			\put(2,45){\makebox(0,0){\scriptsize\rotatebox{90}{Power transmitted (dBm)}}}
			
			\put(52,32){\makebox(0,0){\scriptsize Frequency (GHz)}}
			\put(35.5,45){\makebox(0,0){\scriptsize\rotatebox{90}{Power transmitted (dBm)}}}
			
			\put(85,32){\makebox(0,0){\scriptsize Frequency (GHz)}}
			\put(68.5,45){\makebox(0,0){\scriptsize\rotatebox{90}{Power transmitted (dBm)}}}
			
			\put(18.5,5){\makebox(0,0){\scriptsize Frequency (GHz)}}
			\put(2,18){\makebox(0,0){\scriptsize\rotatebox{90}{Power received (dBm)}}}
			
			\put(52,5){\makebox(0,0){\scriptsize Frequency (GHz)}}
			\put(35.5,18){\makebox(0,0){\scriptsize\rotatebox{90}{Power received (dBm)}}}
			\put(43.5,19.5){\makebox(0,0){\rotatebox{0}{\scriptsize\color{black}\shortstack{\textsc{Peak due to}\\\textsc{LWA 2}}}}}
			\put(58,19.5){\makebox(0,0){\rotatebox{0}{\scriptsize\color{black}\shortstack{\textsc{Peak due to}\\\textsc{LWA 1}}}}}
			\put(85,5){\makebox(0,0){\scriptsize Frequency (GHz)}}
			\put(68.5,18){\makebox(0,0){\scriptsize\rotatebox{90}{Power received (dBm)}}}
		\end{overpic}
	\end{subfigure}

	\caption{LWA-to-LWA analytical demonstration using Friis transmission equation for three cases: a) single LWA-to-LWA (using LWA \#1, broadside at $60$~GHz) with a transmit power $0$~dBm across all frequencies, b) two LWA-to-LWA (using LWA \#1 \& LWA \#2, broadside frequencies at $58$ and $60$~GHz) with the transmit power to each antenna is a channel centered at the broadside frequency, c) $N=4$~LWA-to-LWA (using LWA \#1 to LWA \#4, broadside frequencies at $58$, $60$, $62$, and $64$~GHz) with the transmit power to each antenna is a channel centered at the broadside frequency. The power transmitted and received at each LWA port are shown in each case.}
	\label{fig:LWA_LWA_Friis}
\end{figure*}

To further illustrate the proposed concept of Fig.~\ref{fig:LWA_Proposed}, Friis transmission equation \cite{Johnson} can be used to analytically describe the multiplexing/demultiplexing operation using a multi-port LWA. At a certain frequency, $f$, Friis transmission equation can be written as:

\begin{equation}
\label{Friis}
\frac{P_r }{P_t}= \frac{\lambda^2 G_t(\theta_t,\phi_t)G_r(\theta_r,\phi_r)}{(4\pi R)^2},
\end{equation}

\noindent where $P_t$ is the power input to the transmit antenna, $P_r$ is the power output at the receive antenna, $G_t$ and $G_r$ are the gains of the transmit and receive antennas at frequency, $f$, $R$ is the distance between the antennas, and $\lambda$ is the wavelength of $f$. Eq.~\ref{Friis} assumes the transmit and receive antennas are placed in each other's far-field, perfectly efficient and matched at their ports, and exhibit no polarization loss.

First, a horn-to-LWA transmission calculation will be used to demonstrate demultiplexing of an incident wideband plane wave into unique channels at the LWA outputs. Second, a LWA-to-LWA transmission calculation will be used to demonstrate channel inputs to transmit LWAs (multiplexing) and the extraction of the same channels at the receiving LWAs (demultiplexing). For simplicity and convenience of mathematical description, Gaussian curves are used to model radiation patterns of LWAs~\cite{Emara_DF, ELINT}, i.e.

\begin{equation}\label{Eq:Gaussian}
G(\theta) = A^2 e^{\left[\frac{-k^2(\theta-\alpha)^2}{\theta_B^2}\right]},
\end{equation}

\noindent where $A$ is the voltage gain ($A=1$ assumed here), $\theta_B$ is the $3$-dB beamwidth, $\alpha$ is the squint angle from $\theta=0\degree$, and $k^2 = (2 \ln 4)$ such that $G(\theta_B) = 0.5$ (i.e. $-3$~dB). Let us consider now the two different cases of Fig.~\ref{fig:LWA_Proposed}, separately.\\

\subsubsection{Horn-to-LWA Communication}

Consider a broadband horn antenna used as a transmitter (Tx) exhibiting a frequency independent radiation pattern modeled by the Gaussian profile of \eqref{Eq:Gaussian} and with a peak gain of 0~dB, for simplicity, as shown in Fig.~\ref{fig:horn_LWA_Friis}(a). Next also consider a LWA acting as a receiver (Rx) with a broadside radiation frequency of 60~GHz exhibiting a linear frequency scanning rate of $2.5\degree$/GHz and constant beamwidth of $4\degree$, as shown in Fig.~\ref{fig:horn_LWA_Friis}(b) for Port 1 and Port 2 excitations. The horn antenna and the LWA form an LOS Tx-Rx link with a separation of 2~m and a transmit power of $P_t = 0$~dBm at all frequencies of interest. We would now like to compute the power received by each port of the LWA following the Friis equation of \eqref{Friis}.

Let us first consider the horn antenna oriented at broadside with respect to LWA \#1 as shown in Fig.~\ref{fig:horn_LWA_Friis}(d). The corresponding power $P_r$ versus frequency received at the two ports are also shown. As can be seen, both LWA ports receive the same power profile centered at 60 GHz since it corresponds to the broadside radiation of the LWA. The power received at neighboring frequencies on the other hand, is gradually rejected following the frequency dependent patterns of Fig.~\ref{fig:horn_LWA_Friis}(b). There is no frequency discrimination at the two ports. Let us now tilt the horn antenna so that the incoming radiation to the LWA is off-broadside ($\theta=10\degree$), as shown in Fig.~\ref{fig:horn_LWA_Friis}(e). The received power $P_r$ versus frequency is also shown in Fig.~\ref{fig:horn_LWA_Friis}(e) where in this case, the two ports receive power profiles peaking at \emph{different frequencies}. Port 1 receives peak power at $64$~GHz, corresponding to the forward frequency at $10\degree$ as shown in Fig.~\ref{fig:horn_LWA_Friis}(b). Similarly, Port 2 receives peak power at $56$~GHz, corresponding to the backward frequency at $10\degree$ as shown in Fig.~\ref{fig:horn_LWA_Friis}(b). The LWA thus discriminated the two frequencies following its beam-scanning profile.

Let us next introduce a second LWA which is designed with a different beam-scanning law with 57 GHz at broadside, as compared to that of 60 GHz of the first LWA. Hence 60 GHz is now $+7.5\degree$ for the first port and $-7.5\degree$ for the second port of the LWA, as shown in Fig.~\ref{fig:horn_LWA_Friis}(c). Now, we setup a third experiment where these two LWAs are placed side-by-side forming a 4-port structure, which is receiving a broadband signal from a horn antenna oriented at $+10\degree$, as illustrated in Fig.~\ref{fig:horn_LWA_Friis}(f). As expected, Port 1 and Port 2 of LWA \#1 receive the same power as the previous case in Fig.~\ref{fig:horn_LWA_Friis}(e). Port 3 and Port 4 of LWA \#2 receive peak power at $53$ \& $61$~GHz, corresponding to the forward and backward frequencies specific to LWA 2 at $10\degree$. Therefore, this 4-port antenna structure successfully discriminated 4 frequencies of the incoming broadband signal, effectively demultiplexing them at the 4 ports. This analysis can naturally now be extended to $N$ LWAs for demultiplexing $2N$ frequencies.

\subsubsection{LWA-to-LWA Multiplexing/Demultiplexing}
	
	In this demonstration, the goal is to transmit $N$ channels using $N$ LWAs (Tx), and then receive the same $N$ channels using identical $N$ LWAs (Rx) placed in the far-field, as illustrated in Fig.~\ref{fig:LWA_Proposed}(b). For this demonstration, four pairs of LWAs with different beam-scanning laws will be used such that the broadside frequencies are at $60$ GHz (LWA \#1), $58$ GHz (LWA \#2), $62$ GHz (LWA \#3), and $64$ GHz (LWA \#4). All LWAs have a $2.5\degree$/GHz scanning from the backward to the forward regions, and the $3$-dB beamwidth of $4\degree$ at all frequencies. The separation between the antennas in all cases is $R = 2$~m.
	
	Let us begin with a simple case of a single LWA \#1 pair communicating in an LOS as shown in Fig.~\ref{fig:LWA_LWA_Friis}(a). This first case would serve to demonstrate the frequency selective nature of the LWA at broadside when a constant $P_t=0$~dBm power is transmitted from LWA \#1 (top) to LWA \#1 (bottom) as shown in Fig.~\ref{fig:LWA_LWA_Friis}(a). Since the two antennas are identical and oriented along broadside of each other, the Friis transmission equation is given by
	\begin{equation}
		P_r(f)=   \left[\frac{\lambda G(\theta = 0\degree, f)}{4\pi R}\right]^2,
	\end{equation}
	\noindent where $\theta=0\degree$ corresponds to the broadside direction. The received power $P_r$ by the Rx LWA is also shown in Fig.~\ref{fig:LWA_LWA_Friis}(a), which is seen to be is peaked at $60$~GHz, corresponding to the broadside frequency of LWA \#1. The bandwidth of this channel is simply determined by the beam-scanning law (i.e. $2.5\degree$/GHz and the $4\degree$ $3$-dB beamwidth).
	
	Next consider two different LWA pairs, \#1 and \#2, as shown in Fig.~\ref{fig:LWA_LWA_Friis}(b) where each port of the Tx LWA is excited, i.e. Tx \#1 and Tx \#3. The transmitted power ($P_t$) to each antenna will be a channel centered at the broadside frequency of the antenna with peak power $P_{t,\text{max}}=0$~dBm, centered at $58$ \& $60$~GHz, respectively, as shown in Fig.~\ref{fig:LWA_LWA_Friis}(b). Each channel is emulated using a Gaussian frequency profile. The channels' $3$-dB bandwidths were chosen to be the same as the bandwidth received from the wideband signal shown in Fig.~\ref{fig:LWA_LWA_Friis}(a). The two channels are received using two LWAs identical to the Tx antennas (bottom) through port Rx \#1 (paired with port Tx \#1) and port Rx \#3 (paired with port Tx \#3). Each of the Rx LWA, receives not only signal through its paired Tx, but also from the other transmitting Tx antenna. Consequently, the power received at each frequency by the $n^\text{th}$ antenna is given in terms of $m$ transmitters as, 
	\begin{equation}\label{Eq:FrissLWA_LWA}
		P_{r,n}(f)=   \left[\frac{\lambda}{4\pi R}\right]^2 G_n(\theta = 0\degree, f)  \sum_m G_m(\theta = 0\degree, f)
	\end{equation}
	Using this power equation, the power received by each antenna is computed and is shown in Fig.~\ref{fig:LWA_LWA_Friis}(b).	As expected, Rx port \#1 dominantly receives the channel transmitted by Tx port \#1, and Rx port \#3 dominantly receives a different frequency channel transmitted by Tx port \#3. However, due to coupling between cross Tx and Rx ports accounted by the second term of \eqref{Eq:FrissLWA_LWA}, small power peaks are also observed in the received signal profile at each Rx antenna ($\approx 12$~dB below the main peak). The number of channels can now be increased by introducing more number of unique LWA pairs, as illustrated in Fig.~\ref{fig:LWA_LWA_Friis}(c), where another example is shown for successfully communicating 4 channels centered at $58$, $60$, $62$, and $64$~GHz, using 4 LWA pairs.

\begin{figure}[!b]
	\centering
		\begin{subfigure}{0.5\textwidth}
			\centering
			\begin{overpic}[grid=false, scale = 0.5]{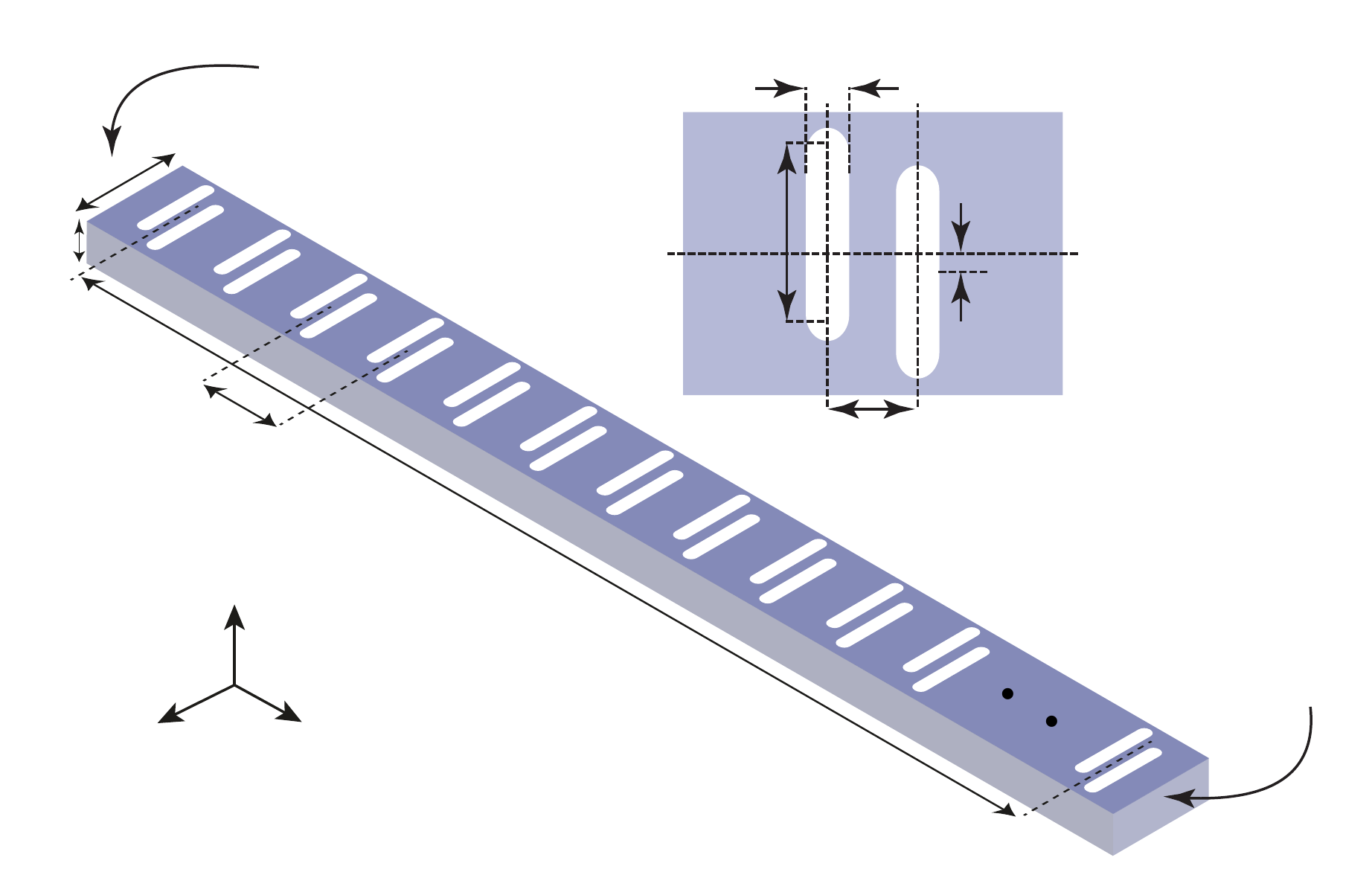}
				\put(10,10){\makebox(0,0){\scriptsize $x$}}		
				\put(22.7,10){\makebox(0,0){\scriptsize $y$}}			
				\put(17,20){\makebox(0,0){\scriptsize $z$}}
				
				\put(64,30.5){\makebox(0,0){\scriptsize $D_s$}}
				\put(60.5,59){\makebox(0,0){\scriptsize $w_s$}}
				\put(56.3,46){\makebox(0,0){\scriptsize $l_s$}}
				\put(70,50){\makebox(0,0){\scriptsize $D_x$}}
				
				\put(37,23){\makebox(0,0){\rotatebox{-30}{\scriptsize $l = M\times p$}}}
				\put(15.5,32.5){\makebox(0,0){\rotatebox{-30}{\scriptsize $p$}}}
				
				\put(4.5,46){\makebox(0,0){\scriptsize $h$}}
				\put(14,53){\makebox(0,0){\scriptsize $w_x$}}
				
				\put(20,58){{\scriptsize Port 1 (left)}}
				\put(94.5,12){\rotatebox{90}{{\scriptsize Port 2 (right)}}}
				
			\end{overpic}
			\caption{}
		\end{subfigure}
	
		\begin{subfigure}{0.25\textwidth}
		\centering
		\begin{overpic}[grid=false, scale = 0.32]{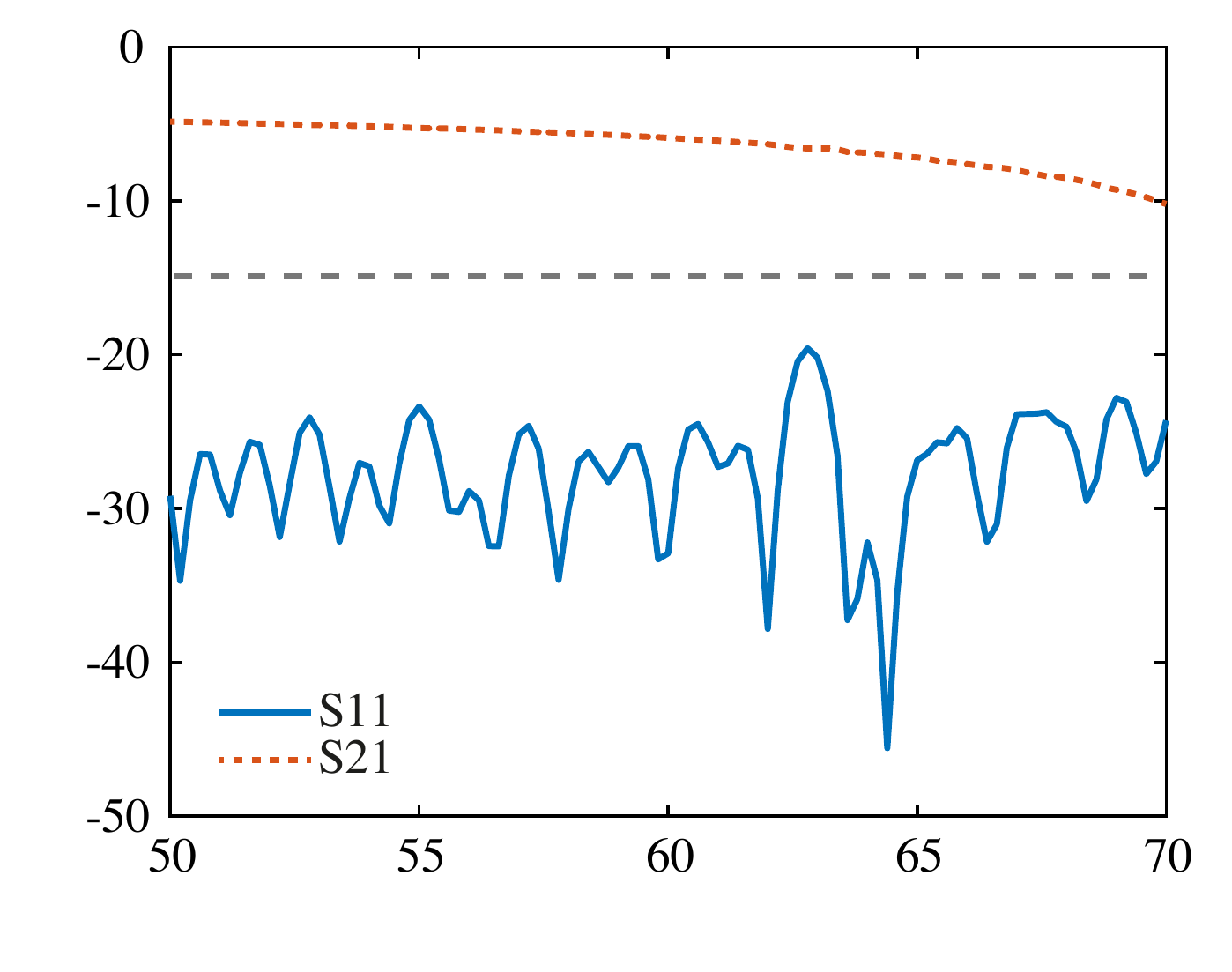}
			\put(54,4.5){\makebox(0,0){\scriptsize Frequency (GHz)}}
			\put(5,45){\makebox(0,0){\scriptsize\rotatebox{90}{Reflection/Transmission (dB)}}}
			\put(44,53){\makebox(0,0){\scriptsize\shortstack{\textsc{Closed Stopband}}}}
		\end{overpic}
		\caption{}
		\end{subfigure}%
		\begin{subfigure}{0.25\textwidth}
		\centering
		\begin{overpic}[grid=false, scale = 0.32]{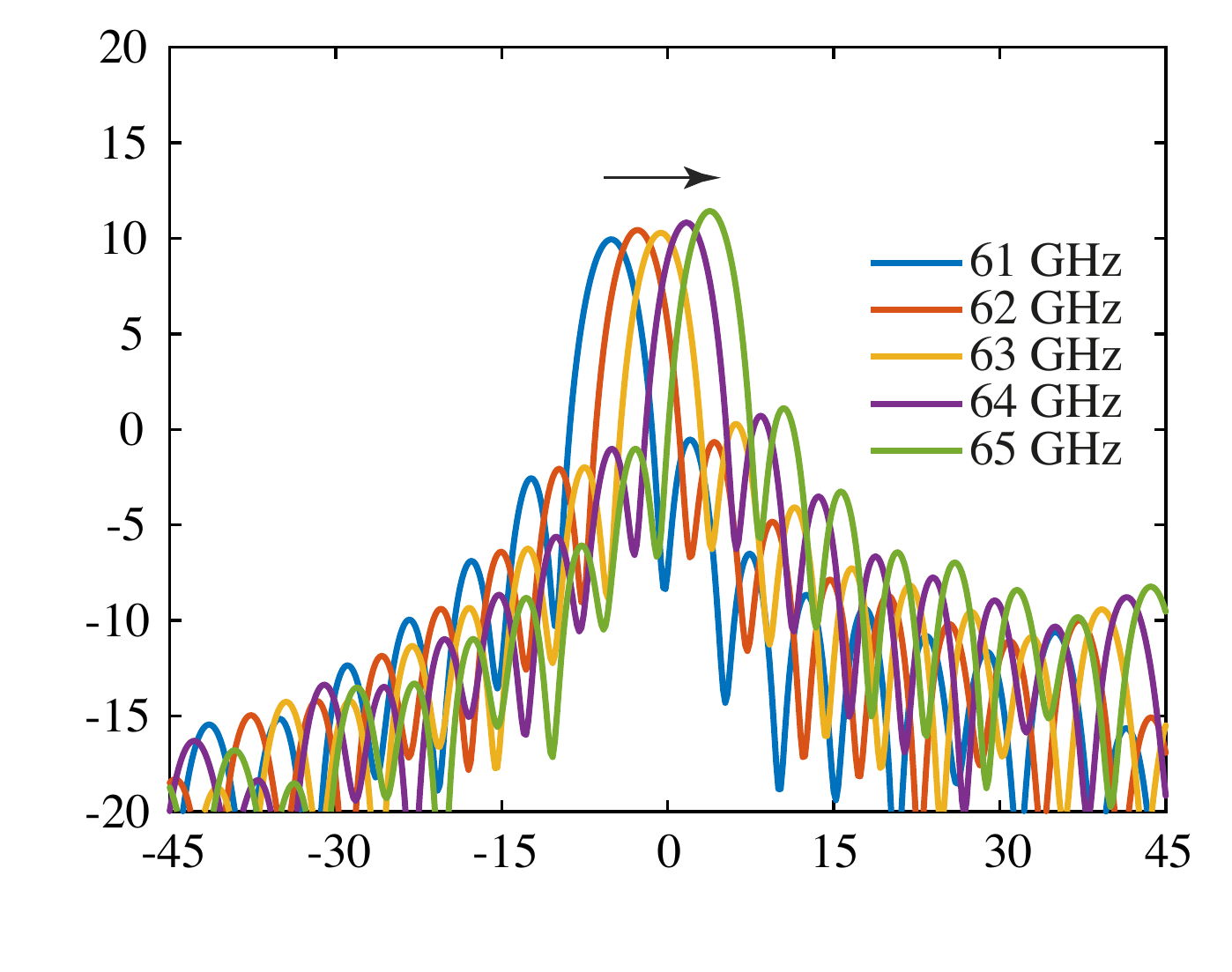}
			\put(54,4.5){\makebox(0,0){\scriptsize Theta, $\theta(\degree)$)}}
			\put(5,45){\makebox(0,0){\scriptsize\rotatebox{90}{Gain (dBi)}}}
			\put(54,67){\makebox(0,0){\scriptsize\shortstack{\textsc{Increasing Frequency}}}}
		\end{overpic}
		\caption{}
		\end{subfigure}
	
		\begin{subfigure}{0.25\textwidth}
		\centering
		\begin{overpic}[grid=false, scale = 0.32]{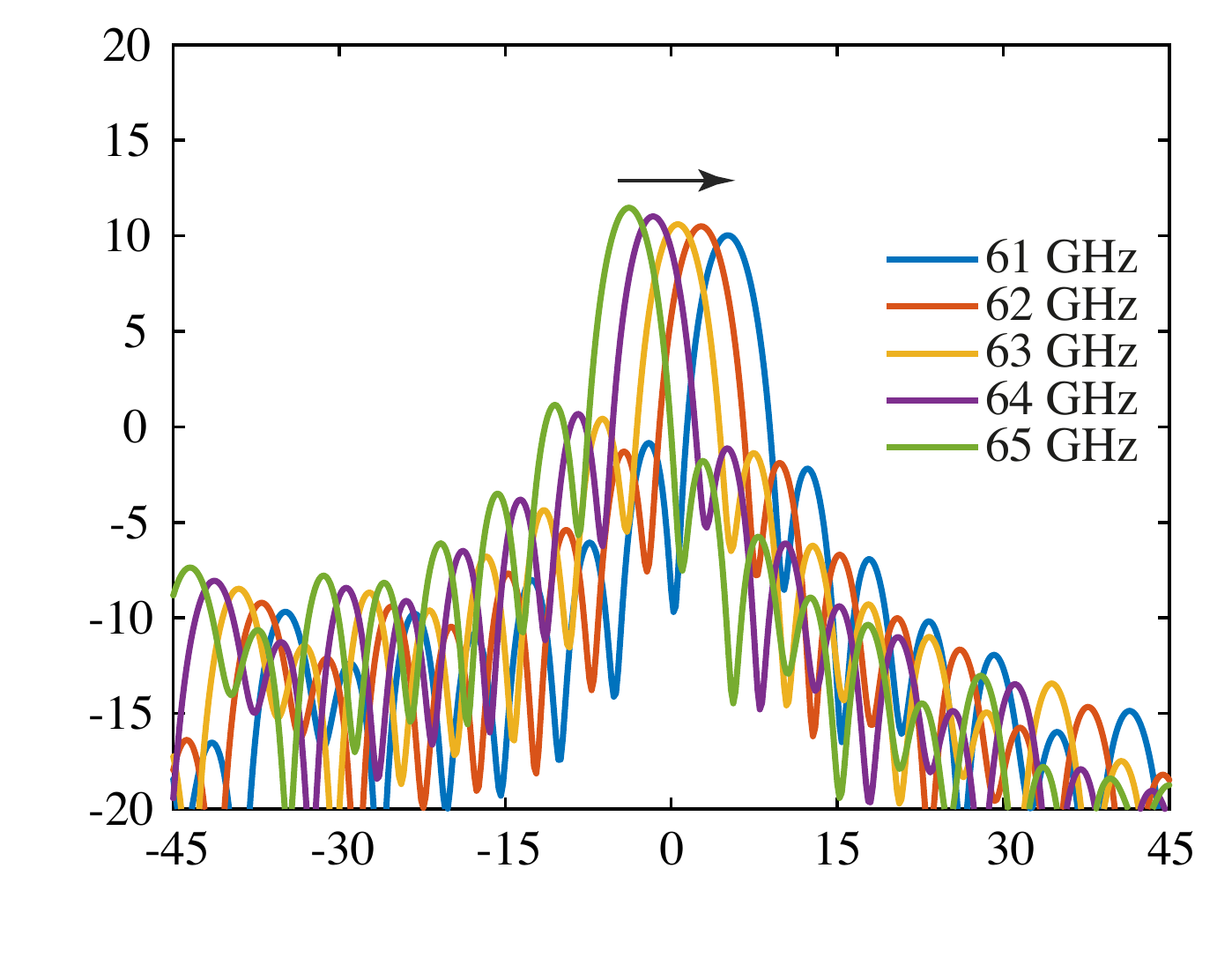}
			\put(54,4.5){\makebox(0,0){\scriptsize Theta, $\theta(\degree)$)}}
			\put(5,45){\makebox(0,0){\scriptsize\rotatebox{90}{Gain (dBi)}}}
			\put(54,67){\makebox(0,0){\scriptsize\shortstack{\textsc{Decreasing Frequency}}}}
		\end{overpic}
		\caption{}
		\end{subfigure}%
		\begin{subfigure}{0.25\textwidth}
		\centering
		\begin{overpic}[grid=false, scale = 0.32]{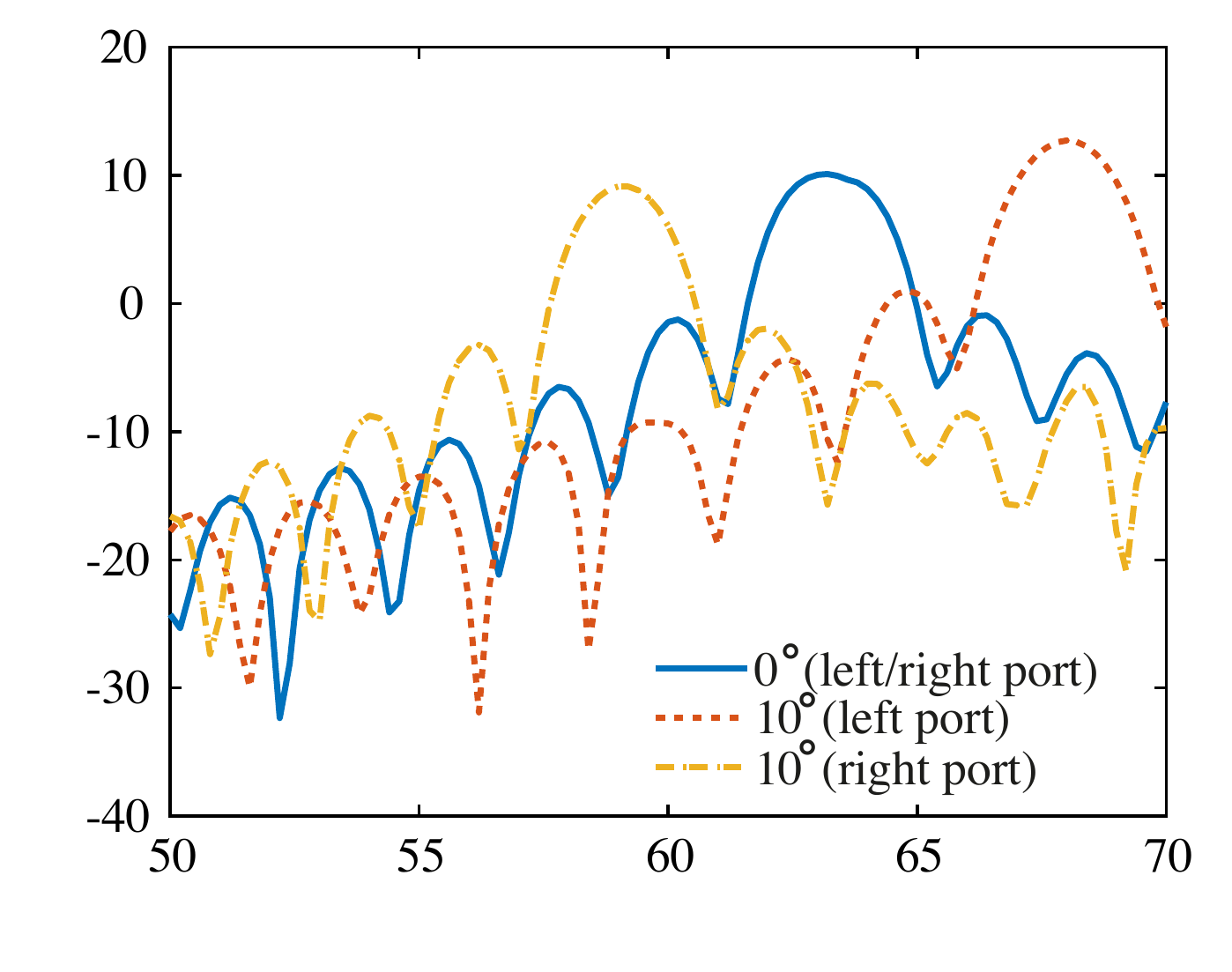}
			\put(54,4.5){\makebox(0,0){\scriptsize Frequency (GHz)}}
			\put(5,45){\makebox(0,0){\scriptsize\rotatebox{90}{Gain (dBi)}}}
		\end{overpic}
		\caption{}
		\end{subfigure}

	\caption{Reflection-cancelling slot pair LWA implemented using a substrate integrated waveguide (SIW) a) Schematic of the antenna implemented on Rogers RO4350 ($\epsilon_r = 3.66$, $\tan\delta = 0.004$, $h = 0.762$~mm) with $M=20$ slots and dimensions (all in mm): $p = 2.78$, $w_{x} = 2.5$, $w_{s} = 0.2$, $l_{s} = 1$, $D_{s} = 0.668$, $D_{x} = 0.315$). b) Reflection and transmission simulated in FEM-HFSS, c) radiation patterns from 61-65~GHz with left port excitation, d) radiation patterns from 61-65~GHz with right port excitation, and e) gain versus frequency at $\theta = 0\degree$ and $\theta = 10\degree$. Radiation patterns are co-polarized fields plotted in the $y-z$~plane, i.e. $G_\theta(\theta, \phi = 90\degree)$.}
	
	\label{fig:slot_pair}
\end{figure}

\begin{figure*}[!htbp]
	\centering
		\begin{subfigure}{0.4\textwidth}
			\centering
			\begin{overpic}[grid=false, scale = 0.2]{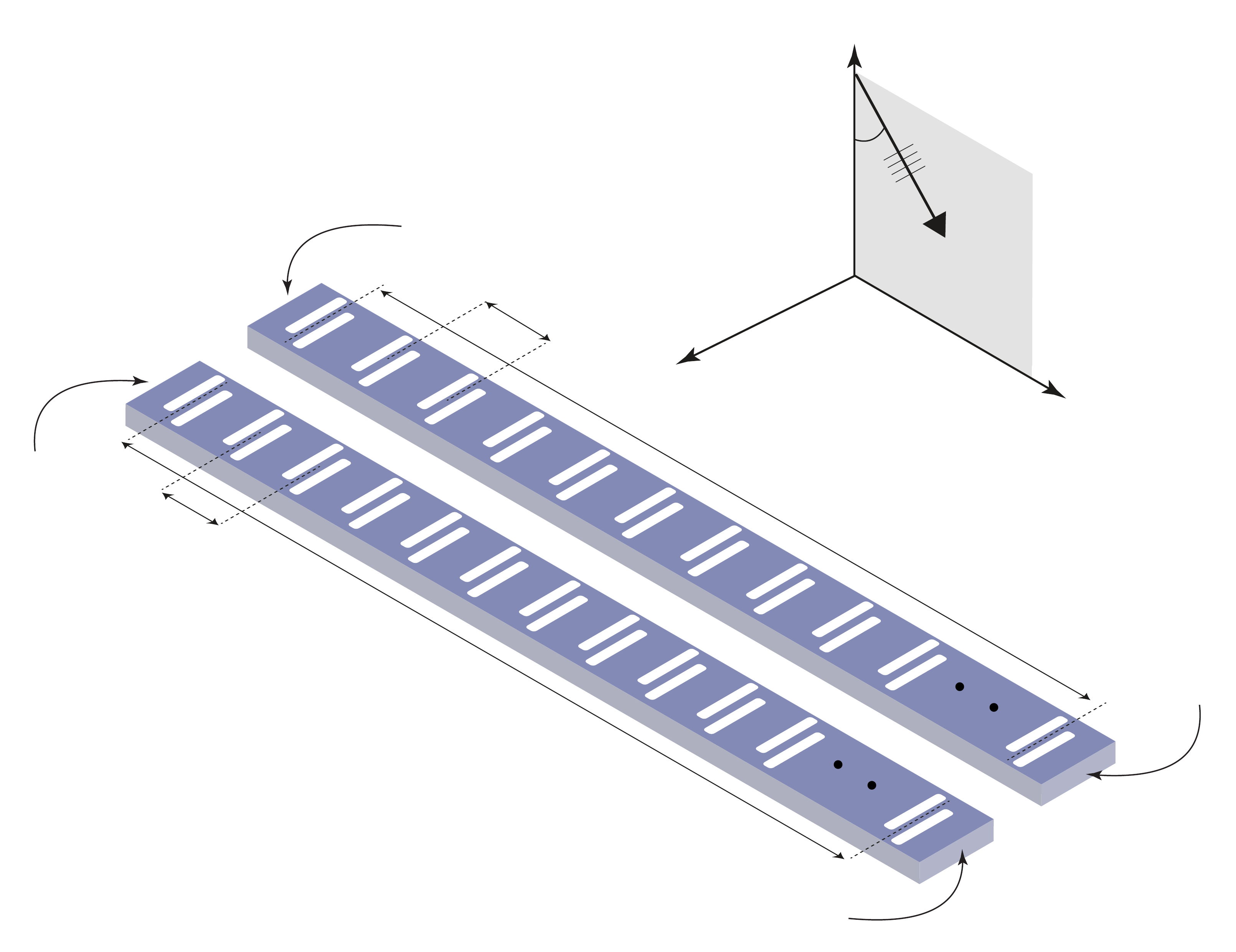}
				\put(53,47){\makebox(0,0){\scriptsize $x$}}		
				\put(87.5,45){\makebox(0,0){\scriptsize $y$}}			
				\put(69,74.5){\makebox(0,0){\scriptsize $z$}}
				\put(71,60){\makebox(0,0){\scriptsize $\theta$}}
				\put(78,57){\makebox(0,0){\scriptsize $k$}}
				
				\put(13.5,34.5){\makebox(0,0){\rotatebox{-30}{\scriptsize $p_1$}}}
				\put(35,23){\makebox(0,0){\rotatebox{-30}{\scriptsize $l_1=M_1\times p_1$}}}
			
				\put(43,52){\makebox(0,0){\rotatebox{-30}{\scriptsize $p_2$}}}
				\put(55.5,41.5){\makebox(0,0){\rotatebox{-30}{\scriptsize $l_2=M_2\times p_2$}}}
				
				\put(53,3){\makebox(0,0){\rotatebox{0}{\scriptsize Ant 1 Port 2 (right)}}}
				\put(96.5,37){\makebox(0,0){\rotatebox{90}{\scriptsize Ant 2 Port 2 (right)}}}
				
				\put(47,58){\makebox(0,0){\rotatebox{0}{\scriptsize Ant 2 Port 1 (left)}}}
				\put(3.5,25){\makebox(0,0){\rotatebox{90}{\scriptsize Ant 1 Port 1 (left)}}}
			\end{overpic}
			\caption{}
		\end{subfigure}%
		\begin{subfigure}{0.3\textwidth}
			\centering
			\begin{overpic}[grid=false, scale = 0.4]{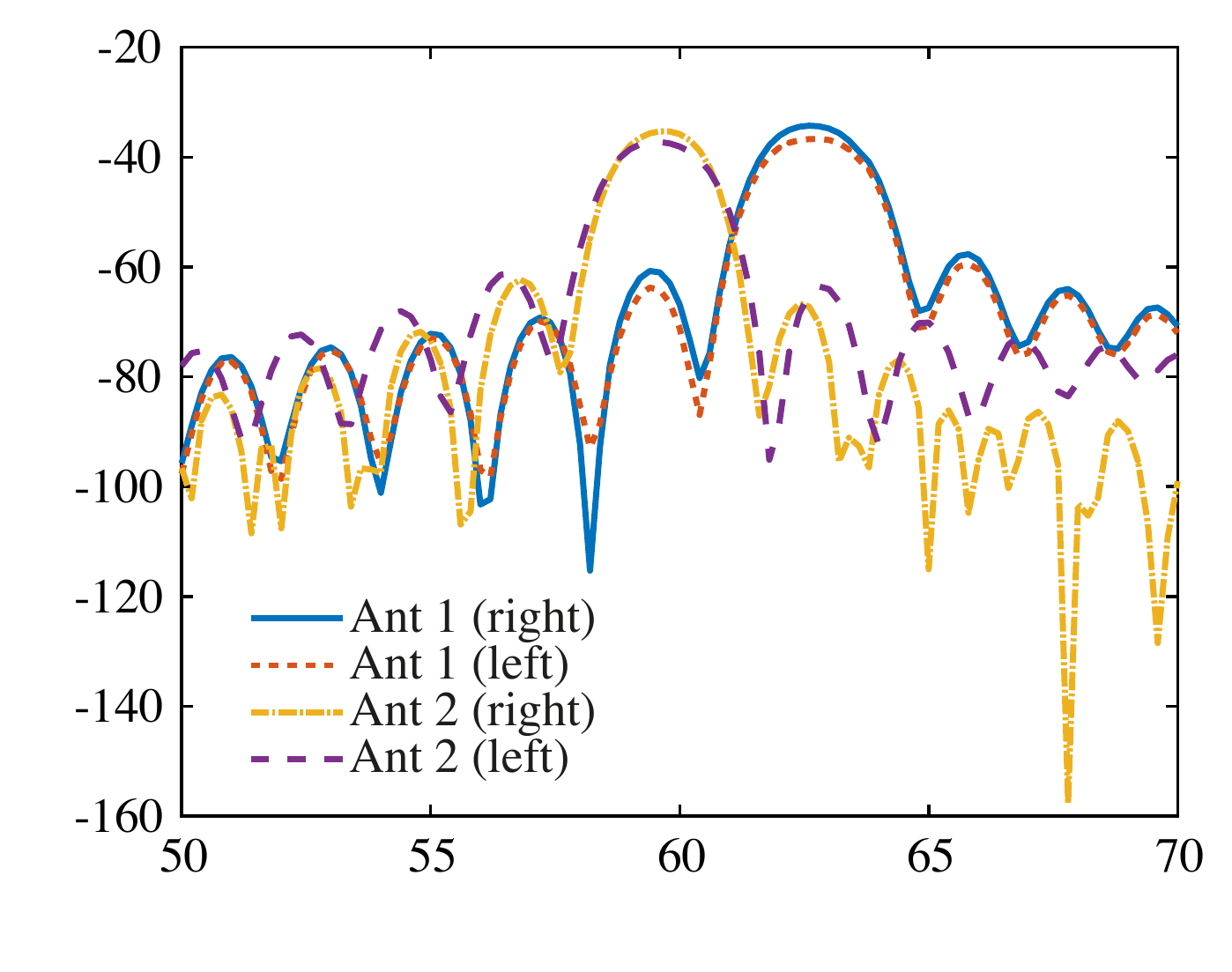}
				\put(54,4.5){\makebox(0,0){\scriptsize Frequency (GHz)}}
				\put(4.5,45){\makebox(0,0){\scriptsize\rotatebox{90}{Power (dBW)}}}
			\end{overpic}
			\caption{}
		\end{subfigure}%
		\begin{subfigure}{0.3\textwidth}
			\centering
			\begin{overpic}[grid=false, scale = 0.4]{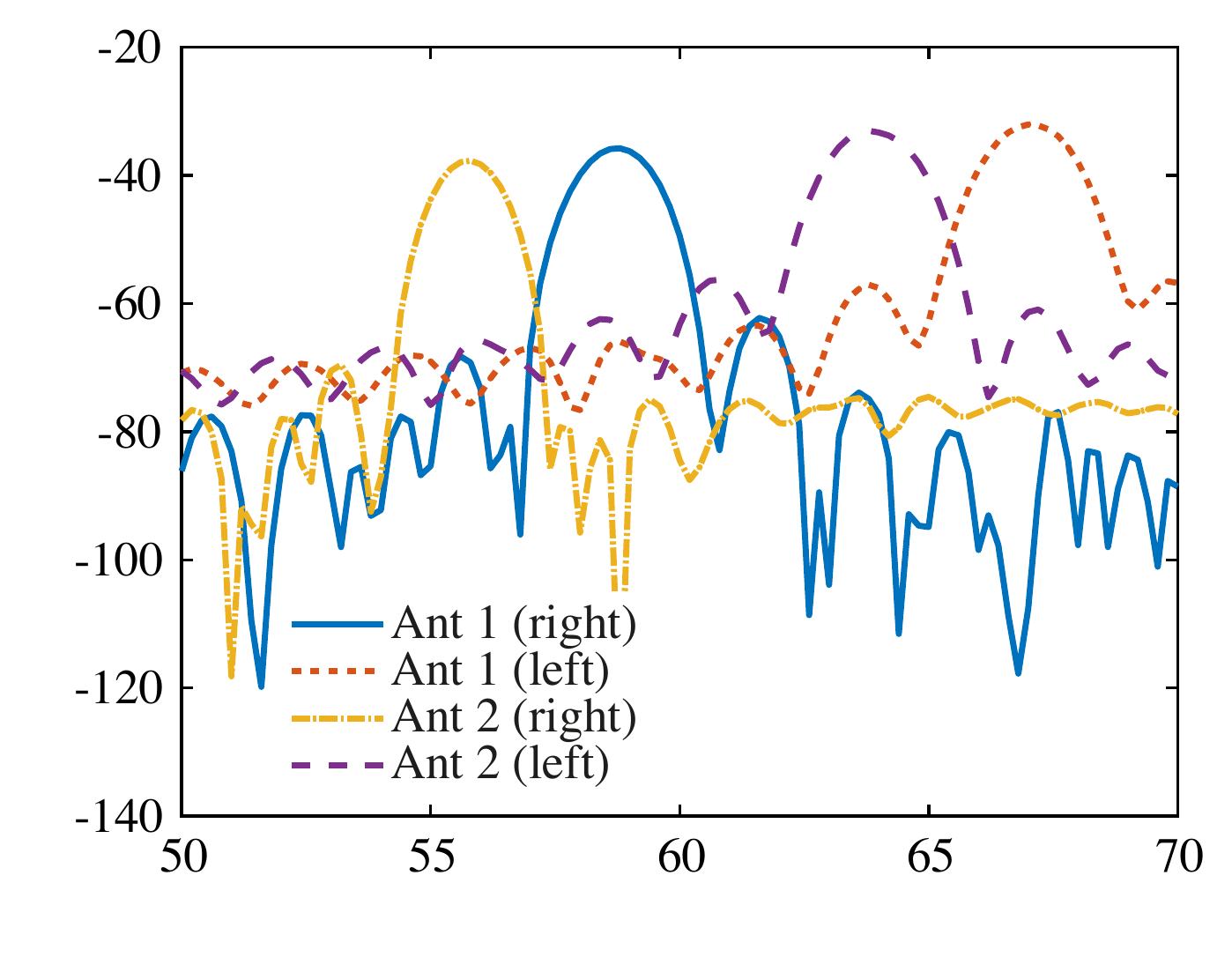}
				\put(54,4.5){\makebox(0,0){\scriptsize Frequency (GHz)}}
				\put(4.5,45){\makebox(0,0){\scriptsize\rotatebox{90}{Power (dBW)}}}
			\end{overpic}
			\caption{}
		\end{subfigure}%

		\begin{subfigure}{1\textwidth}
			\centering
			\begin{overpic}[grid=false, scale = 0.5]{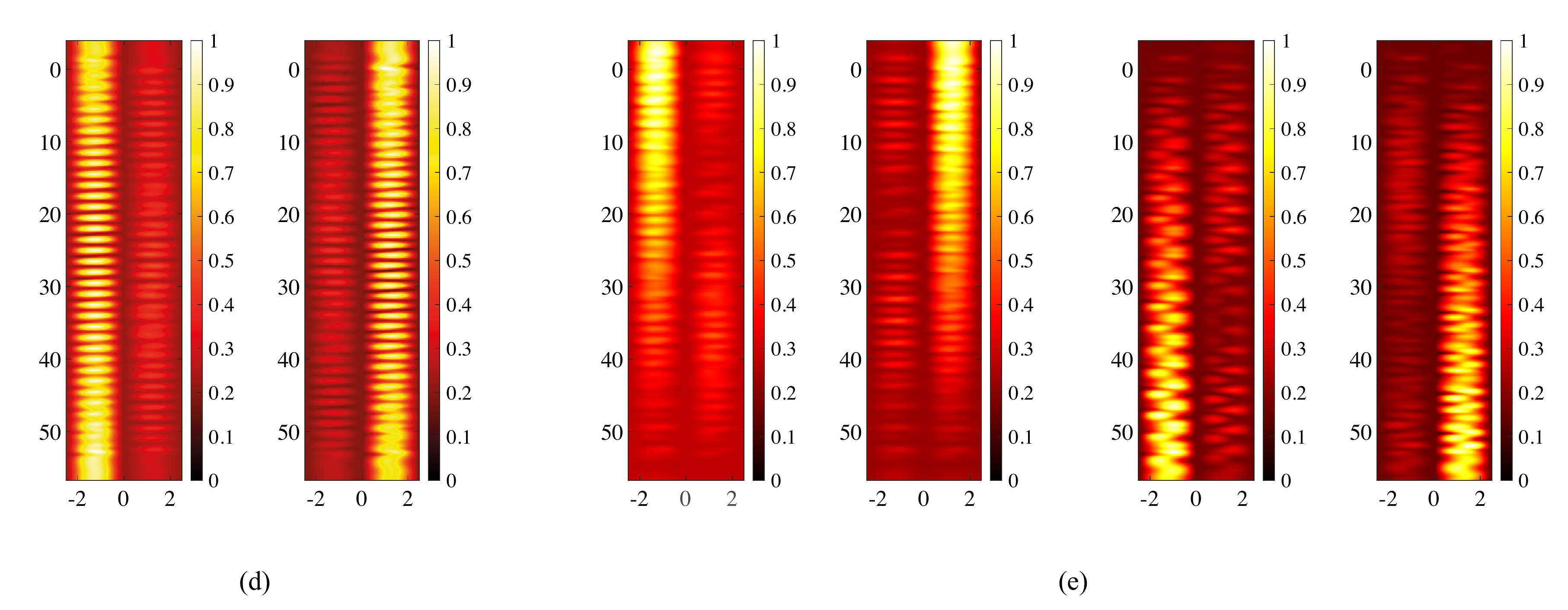}
				\put(8,5.5){\makebox(0,0){\scriptsize $x$~(mm)}}
				\put(1,22.5){\makebox(0,0){\scriptsize\rotatebox{90}{$y$~(mm)}}}
				
				\put(24,5.5){\makebox(0,0){\scriptsize $x$~(mm)}}
				\put(17,22.5){\makebox(0,0){\scriptsize\rotatebox{90}{$y$~(mm)}}}
				
				\put(44,5.5){\makebox(0,0){\scriptsize $x$~(mm)}}
				\put(37,22.5){\makebox(0,0){\scriptsize\rotatebox{90}{$y$~(mm)}}}
				
				\put(59,5.5){\makebox(0,0){\scriptsize $x$~(mm)}}
				\put(52.5,22.5){\makebox(0,0){\scriptsize\rotatebox{90}{$y$~(mm)}}}
				
				\put(76.5,5.5){\makebox(0,0){\scriptsize $x$~(mm)}}
				\put(69.5,22.5){\makebox(0,0){\scriptsize\rotatebox{90}{$y$~(mm)}}}
				
				\put(92,5.5){\makebox(0,0){\scriptsize $x$~(mm)}}
				\put(85,22.5){\makebox(0,0){\scriptsize\rotatebox{90}{$y$~(mm)}}}
				
				\put(8.5,37.5){\makebox(0,0){\color{blue}\scriptsize $f = 59.6$~GHz}}
				\put(24,37.5){\makebox(0,0){\color{blue}\scriptsize $f = 62.6$~GHz}}
				\put(44,37.5){\makebox(0,0){\color{blue}\scriptsize $f = 55.8$~GHz}}
				\put(59,37.5){\makebox(0,0){\color{blue}\scriptsize $f = 58.8$~GHz}}
				\put(77,37.5){\makebox(0,0){\color{blue}\scriptsize $f = 63.8$~GHz}}
				\put(92,37.5){\makebox(0,0){\color{blue}\scriptsize $f = 67$~GHz}}
			\end{overpic}
		\end{subfigure}%

	\caption{Full-wave demonstration of the Horn-to-LWA channel demultiplexer using slot array antennas of Fig.~\ref{fig:slot_pair}. a) Illustration of two slot antennas ($M_1 = 20$ \& $M_2 = 19$) with an incident plane wave in the $yz-$plane at angle, $\theta$, with respect to broadside. The dimensions of each antenna are as follows (all in mm): Antenna 1 ($p_1 = 2.78$, $w_{x1} = 2.5$, $w_{s1} = 0.2$, $l_{s1} = 1$, $D_{s1} = 0.668$, $D_{x1} = 0.315$), and Antenna 2 ($p_2 = 3$, $w_{x2} = 2.5$, $w_{s2} = 0.2$, $l_{s2} = 1$, $D_{s2} = 0.73$, $D_{x2} = 0.315$). Full-wave demonstration results of the power received at all antenna ports with an incident plane wave at a distance of $1$~cm from the antenna and magnitude of $1000$ V/m: b) power received versus frequency with plane wave incident at broadside, $\theta = 0\degree$, c) power received versus frequency with plane wave incident at $\theta = 10\degree$. d) Field distribution at the two broadside peak frequencies: $59.6$ and $62.6$~GHz, and e) field distributions at the $\theta = 10\degree$ four peak frequencies: $55.8$, $58.8$, $63.8$ and $67$~GHz.}
	\label{fig:slot_ant_PW}
\end{figure*}
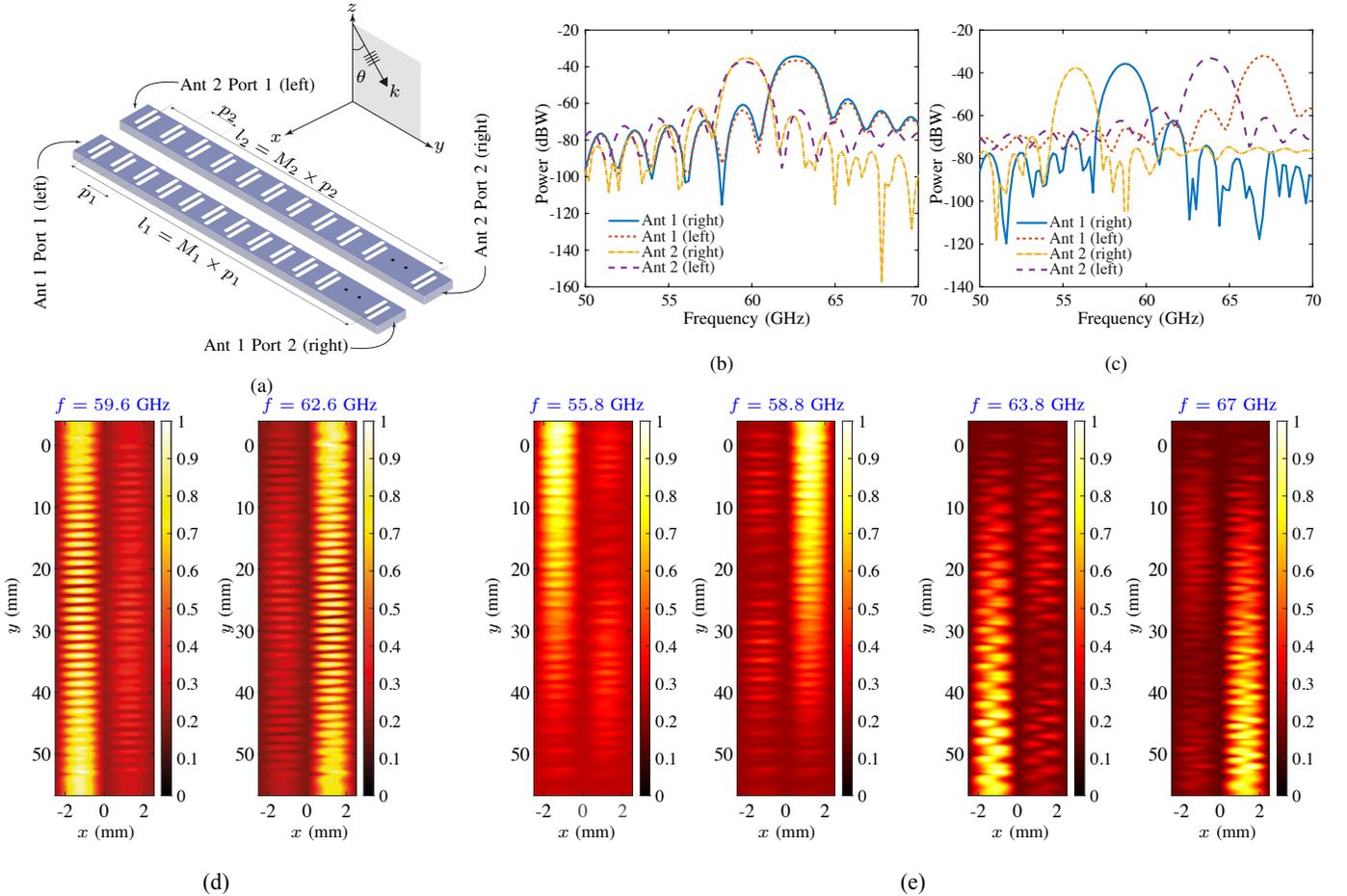

\section{Full-Wave Demonstration using Slot Arrays}

So far, ideal antenna apertures with ideal beam-scanning laws have been considered to demonstrate the proposed demultiplexing and multiplexing principle. In this section, the concept of channel demultiplexing will be demonstrated using a practical slot array antenna. 
\subsection{LWA Implementation Using Reflection-Cancelling Slot Pair LWA}
LWAs can be designed using various techniques such as series-fed patch (SFP) LWAs \cite{Bassam_SFP, Metzler_SFP}, composite right/left-handed (CRLH) LWAs \cite{Caloz_CRLH}, slotted rectangular waveguides \cite{Volakis}, and slotted substrate integrated waveguide (SIW) \cite{Liu_SIW, DJ_SA}. In this paper, LWAs will be implemented using reflection-cancelling slot pairs \cite{Sakakibara_slots}, which are particularly suited for mm-wave application due to their substrate integration and ability to close periodic stop-bands enabling a full-space beam scanning. A pair of offset slots are used as radiating elements, creating the asymmetry needed to suppress the open stopband at broadside \cite{Otto_LWA}. To achieve compactness, simple fabrication, and easy integration with other circuitry, the slot pairs can be implemented on an SIW on a PCB \cite{Deslandes_SIW}. The slot period controls main beam angle by the beam-scanning law \cite{Otto_LWA}

	\begin{equation}
	\theta(\omega) \approx \sin^{-1}\left(\frac{\beta_0+n\pi/p}{k_0}\right),
	\end{equation}
	
	\noindent where $\beta_0$ is the fundamental mode phase constant, $n$ is the radiating space harmonic and it is typically $n = -1$, $p$ is the slot period, and $k_0 = \omega/c$ is the free-space wavenumber. In this equation, the leakage factor is assumed to be negligible in the calculation of the main beam direction \cite{Collin_antenna}.

	The reflection-cancelling slot pairs were next implemented on an SIW as shown in Fig.~\ref{fig:slot_pair}(a). The SIW was implemented on Rogers RO4350 ($\epsilon_r = 3.66$, $\tan\delta = 0.004$, $h = 0.762$~mm). The slot pair dimensions and their relative locations are paramount to close the open stop-band at broadside for high efficiency radiation. The length of the antenna (i.e. number of slot pairs, $M$) determines the leakage which determines the beamwidth. Fig.~\ref{fig:slot_pair}(b) shows the reflection and transmission magnitudes of a typical design. The graph shows low reflection, suggesting that the stop-band is closed. Fig.~\ref{fig:slot_pair}(c) \& (d) show the far-field radiation patterns from $61$ to $65$ GHz for Port 1 (left) and Port 2 (right) excitations, respectively. Fig.~\ref{fig:slot_pair}(e) alternatively shows gain vs. frequency at broadside ($\theta = 0\degree$) and at $\theta = 10\degree$ for Port 1 (left) and Port 2 (right). This practical slot array design will now be used next as a receiver to demultiplex multiple frequency channels.

\subsection{Incident Plane-Wave Simulations}

Consider two different slot array antenna with different broadside frequencies, placed side-by-side, as shown in Fig.~\ref{fig:slot_ant_PW}(a). The slot array period, $p$ is modified in each design to obtain a desired broadside frequency where the channels of interest are centered. In addition, the number of slot arrays, $M$ is chosen to obtain comparable beamwidths for the two antennas. A uniform plane-wave excitation is next incident on the structure at angle of $\theta$ with respect to the aperture normal at a sweeping frequency $f$, and the electromagnetic signal induced at various ports of the two LWAs is recorded. 

Fig.~\ref{fig:slot_ant_PW}(b) shows the power received at all four ports with a normally incident plane-wave along the broadside of the antennas, first. As expected, LWA \#1 Port \#1 (left) and Port \#2  (right) receive the same power profile centered at the broadside frequency of Antenna 1 ($\approx 59.6$~GHz), consistent with the analytical demonstration shown in Fig.~\ref{fig:horn_LWA_Friis}(a). At the same time, LWA \#2 receives peak power corresponding to the broadside frequency of the second LWA ($\approx 60.6$~GHz). Fig.~\ref{fig:slot_ant_PW}(d) further shows the E-field magnitudes at the peak frequencies induced along the LWAs. It clearly shows that only one of the LWAs receives the peak power at its respective frequency, while the other one is effectively isolated. Moreover, as expected under broadside incidence conditions, the induced fields propagate along both ports of the LWAs. Therefore, these two different LWAs act as $1:2$ channel demultiplexer.

In the next case, the incoming plane-wave is tilted to an angle of $\theta=10\degree$. In this case, each of the LWA receives two distinct frequencies corresponding to the forward and backward radiating frequencies along $10\degree$ following their specific beam scanning laws. Fig.~\ref{fig:slot_ant_PW}(c) shows the power received with a plane wave incident at $\theta=10\degree$ at each of the 4 ports. The 4-port antenna structure thus discriminated 4 distinct frequency bands routing them to each of the LWA ports, as clearly observed in the induced E-fields shown in Fig.~\ref{fig:slot_ant_PW}(e), which shows the field distributions at the four peak frequencies of Fig.~\ref{fig:slot_ant_PW}(c). In this case of oblique plane-wave incidence, this LWA pair acts as a $1:4$ demultiplexer. Naturally, being a reciprocal structure, the 4-port antenna structure will act as a $4:1$ multiplexer, radiating all these four frequencies along $\theta= 10\degree$.\footnote{A similar demonstration may also be shown for a LWA-to-LWA communication. However, multiplexing/demultiplexing using Tx and Rx LWAs is not demonstrated here in full-wave analysis due to the large size of the simulation model containing two LWAs in each other's far-field, which makes the simulation time prohibitively long.}

\section{Experimental Demonstration}
\subsection{LWA Characterization}
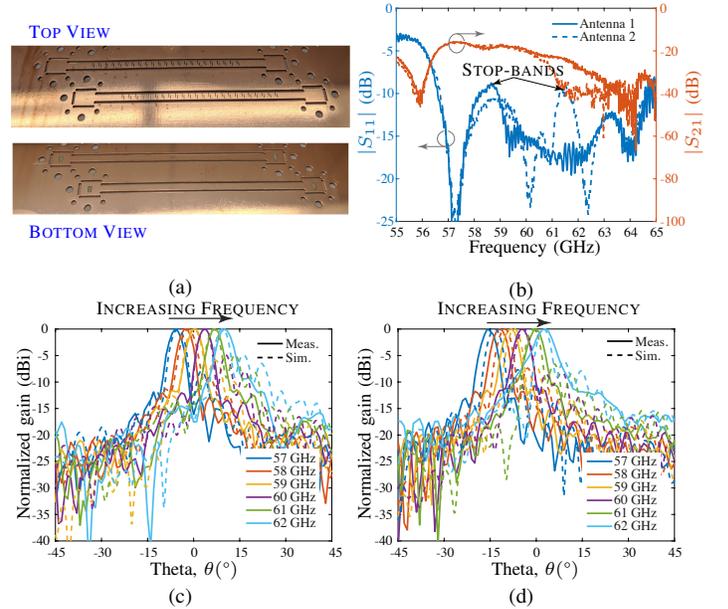
\begin{figure}[!t]
	\centering
	\begin{subfigure}{0.25\textwidth}
			\centering
			\begin{overpic}[grid=false, scale = 0.04]{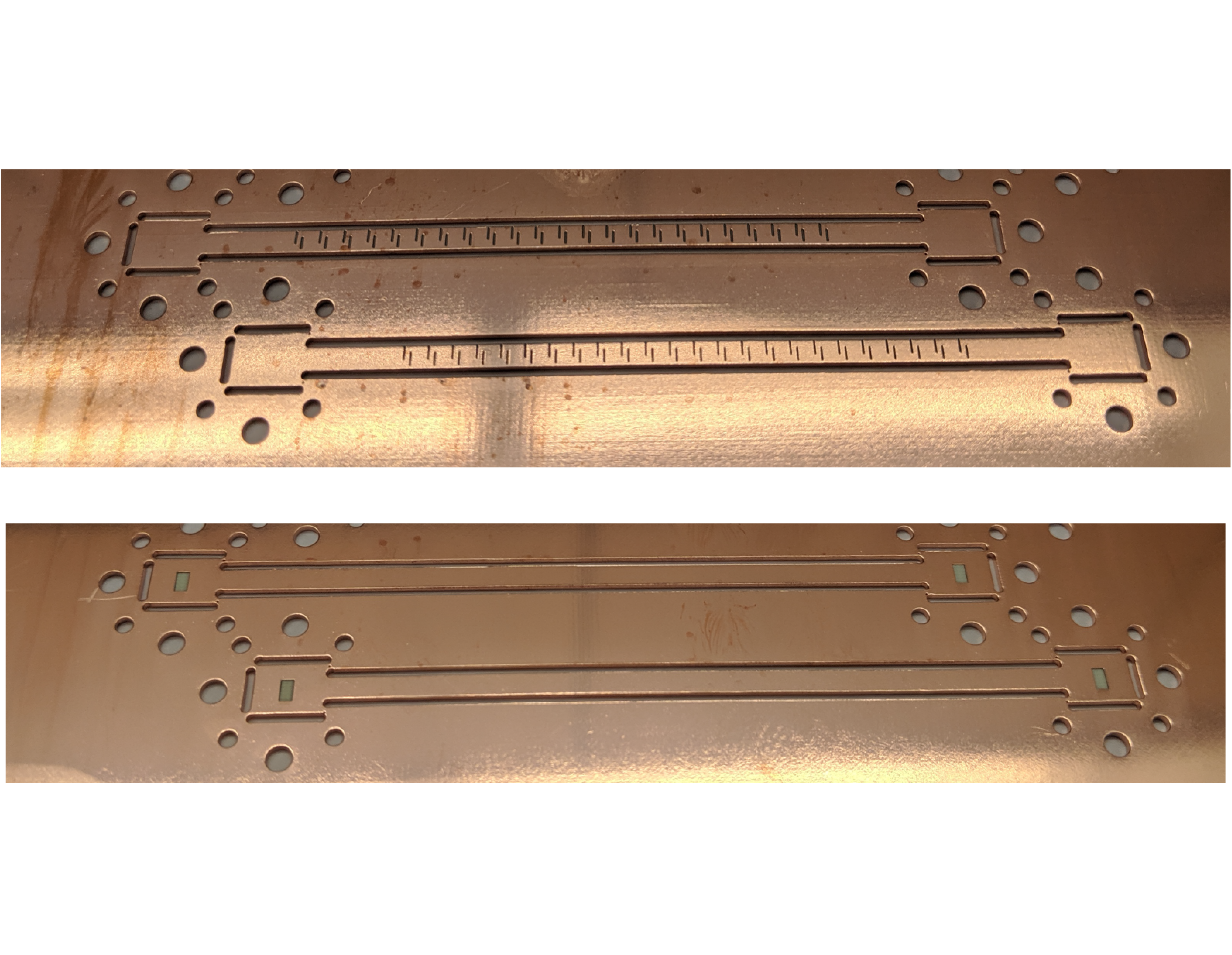}
				\put(5,66){{\color{blue}\scriptsize\shortstack{\textsc{Top View}}}}
				\put(5,7){{\color{blue}\scriptsize\shortstack{\textsc{Bottom View}}}}
			\end{overpic}
			\caption{}
	\end{subfigure}%
	\begin{subfigure}{0.25\textwidth}
		\centering
		\begin{overpic}[grid=false, scale = 0.32]{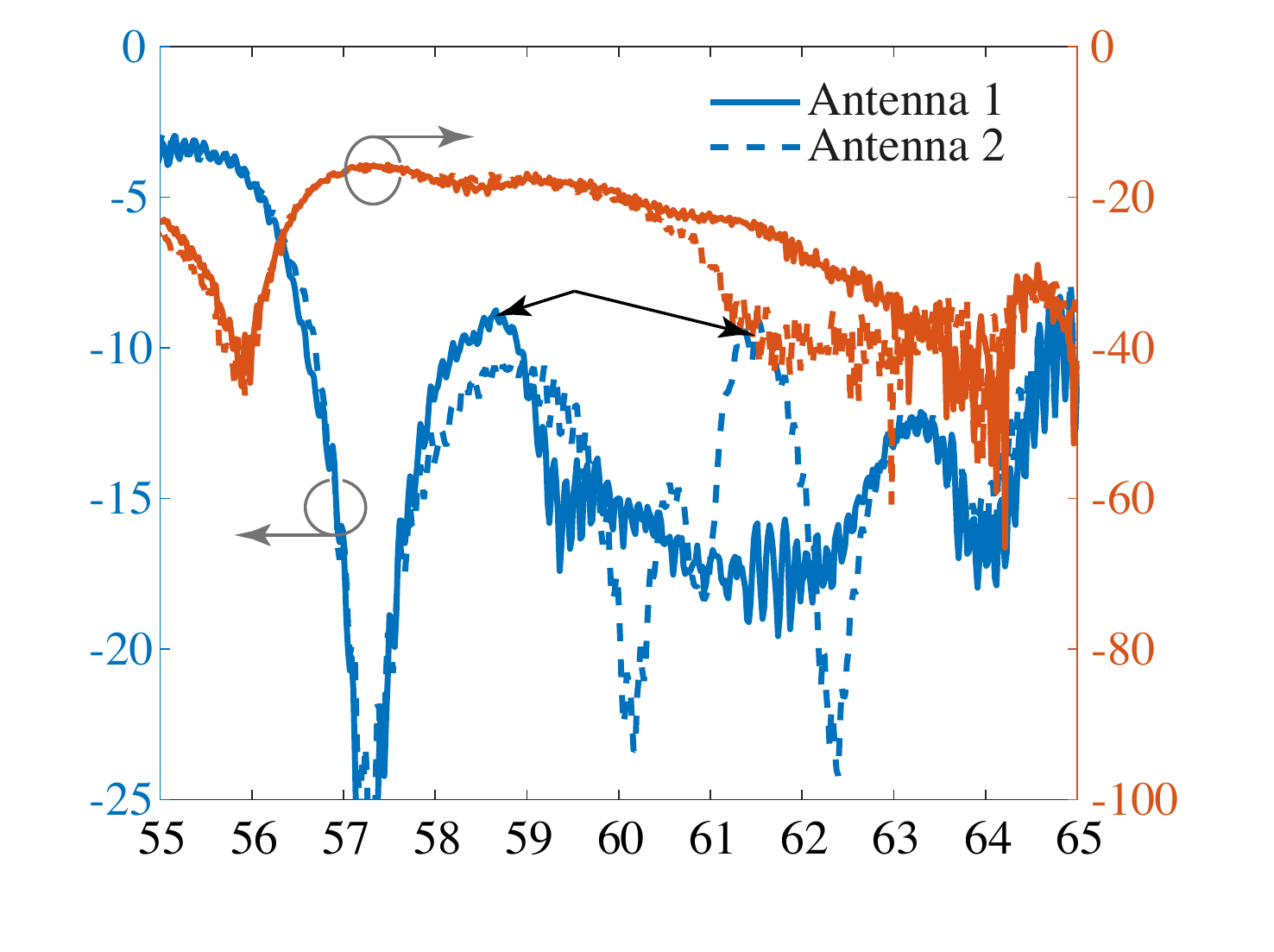}
				\put(51,4.5){\makebox(0,0){\scriptsize Frequency (GHz)}}
				\put(4.5,42){\makebox(0,0){\scriptsize\color{matlabblue}\rotatebox{90}{$|S_{11}|$~(dB)}}}
				\put(95,42){\makebox(0,0){\scriptsize\color{matlaborange}\rotatebox{90}{$|S_{21}|$~(dB)}}}
				\put(45,54){\makebox(0,0){\color{black}\scriptsize\shortstack{\textsc{Stop-bands}}}}
		\end{overpic}
		\caption{}
	\end{subfigure}%

	\begin{subfigure}{0.25\textwidth}
	\centering
		\begin{overpic}[grid=false, scale = 0.32]{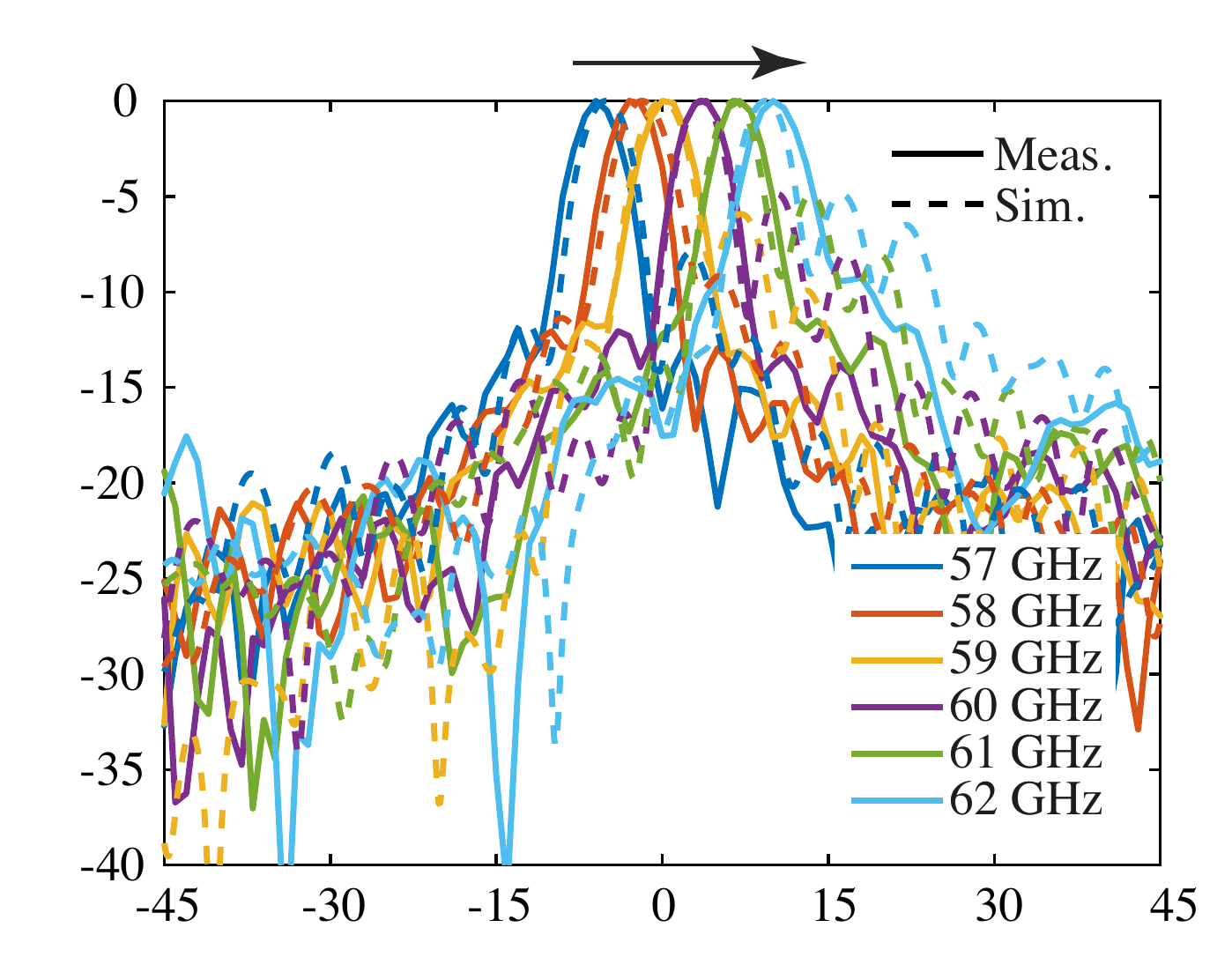}
				\put(54,0){\makebox(0,0){\scriptsize Theta, $\theta(\degree)$}}
				\put(4.5,40){\makebox(0,0){\scriptsize\rotatebox{90}{Normalized gain (dBi)}}}
				\put(55,76.5){\makebox(0,0){\color{black}\scriptsize\shortstack{\textsc{Increasing Frequency}}}}
		\end{overpic}
		\caption{}
	\end{subfigure}%
	\begin{subfigure}{0.25\textwidth}
		\centering
		\begin{overpic}[grid=false, scale = 0.32]{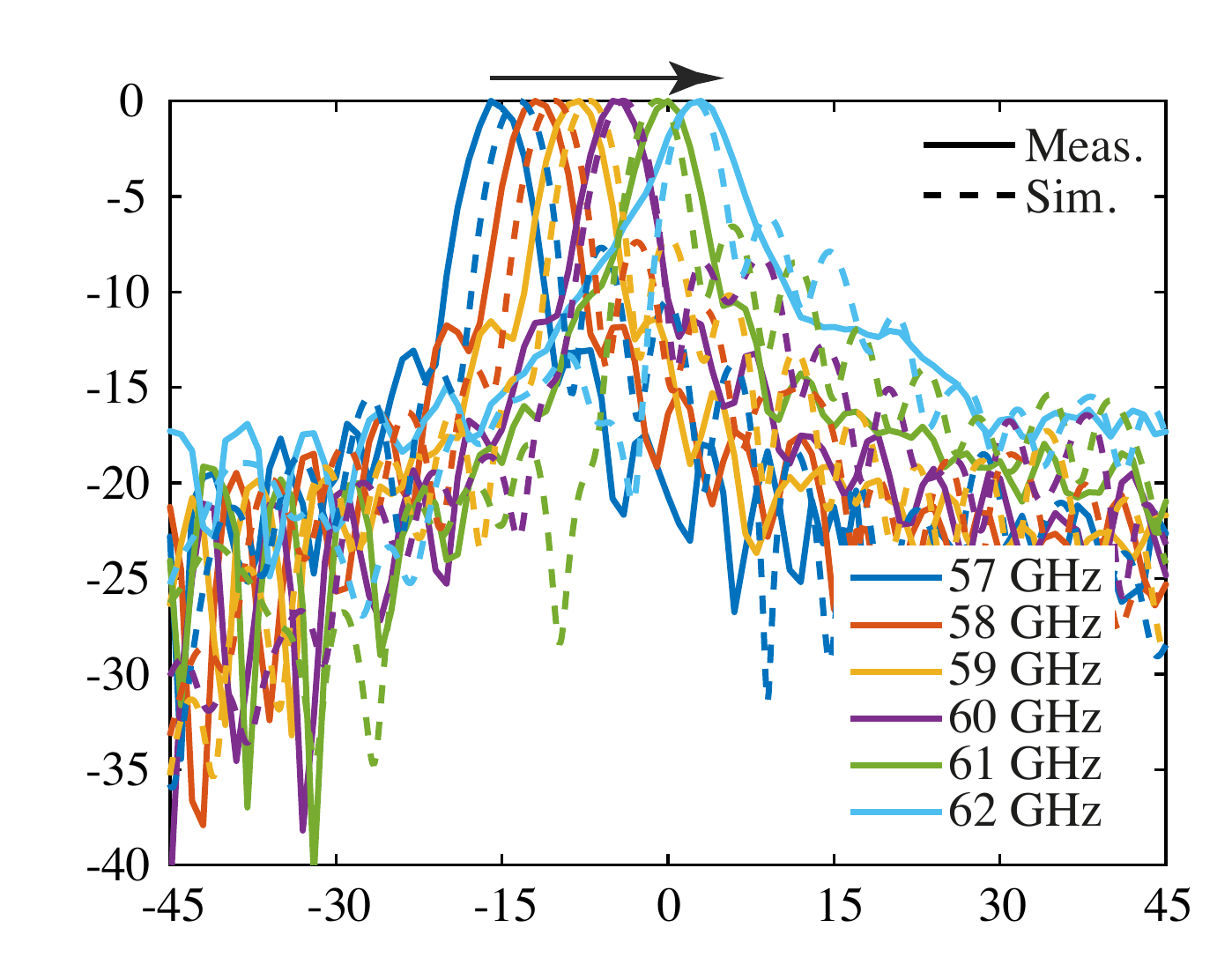}
				\put(54,0){\makebox(0,0){\scriptsize Theta, $\theta(\degree)$}}
				\put(4.5,40){\makebox(0,0){\scriptsize\rotatebox{90}{Normalized gain (dBi)}}}
				\put(55,76.5){\makebox(0,0){\color{black}\scriptsize\shortstack{\textsc{Increasing Frequency}}}}
		\end{overpic}
		\caption{}
	\end{subfigure}%

	\caption{Fabricated slot array antennas. a) Top and bottom views of the fabricated LWAs, b) measured reflection and transmission characteristics of the two antennas, and far-field radiation pattern measurements from $57$ to $62$~GHz of a) antenna 1, and d) antenna 2, with a left port excitation.}
	\label{fig:exp_demo1}
\end{figure}

	Next, two LWAs were fabricated to experimentally demonstrate the proposed applications. The antennas were fabricated on Rogers RO4350 ($\epsilon_r = 3.66$, $\tan\delta = 0.004$, $h = 0.762$~mm) with standard WR-15 waveguide feeds as shown in Fig.~\ref{fig:exp_demo1}(a). The antennas were characterized by performing reflection and far-field anechoic measurements. Reflection results are shown in Fig.~\ref{fig:exp_demo1}(b), showing reflection $<-9$~dB across the bandwidth. While the antennas were designed with a closed stop-band, measurements show a pronounced stop-band at the broadside frequencies for both antennas. This is due to fabrication tolerances to which the stop-band closure is very sensitive, especially to the dimensions of the slot pairs and their relative locations. Fig.~\ref{fig:exp_demo1}(c) and (d) show the simulated and measured far-field radiation patterns from $57$ to $62$~GHz (left port excitation) for antenna 1 and antenna 2, respectively. There is a discrepancy between simulations and measurements which can be attributed to variations in the permittivity and loss tangent between measurements and simulations. Additionally, it was assumed that the permittivity and loss tangent are constant with frequency, which may not be the case in the physical substrate. Finally, a small misalignment during the far-field measurements while mounting the antennas can also cause measurable errors in the radiation patterns.

\subsection{Horn-to-LWA Demultiplexing}

\begin{figure}[!t]
	\begin{minipage}{0.45\columnwidth}
	\begin{subfigure}{\textwidth}
		\begin{overpic}[grid=false, scale = 0.69]{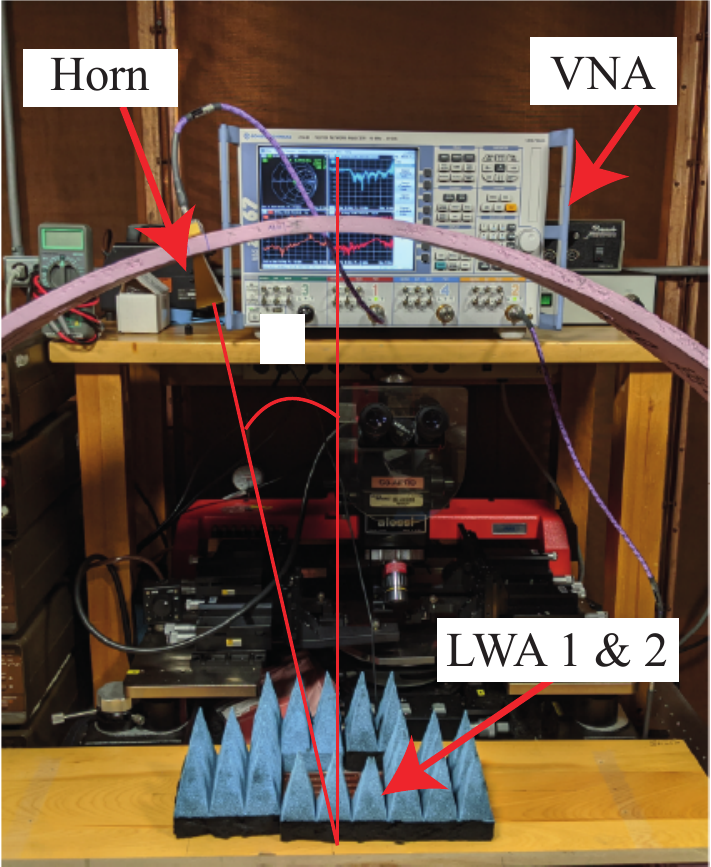}
			\put(32.5,61){\makebox(0,0){\scriptsize $\theta$}}
		\end{overpic}
		\caption{}
	\end{subfigure}%
	\end{minipage}\hspace{1cm}
	\begin{minipage}{0.45\columnwidth}
	\begin{subfigure}{\textwidth}
		\begin{overpic}[grid=false, scale = 0.28]{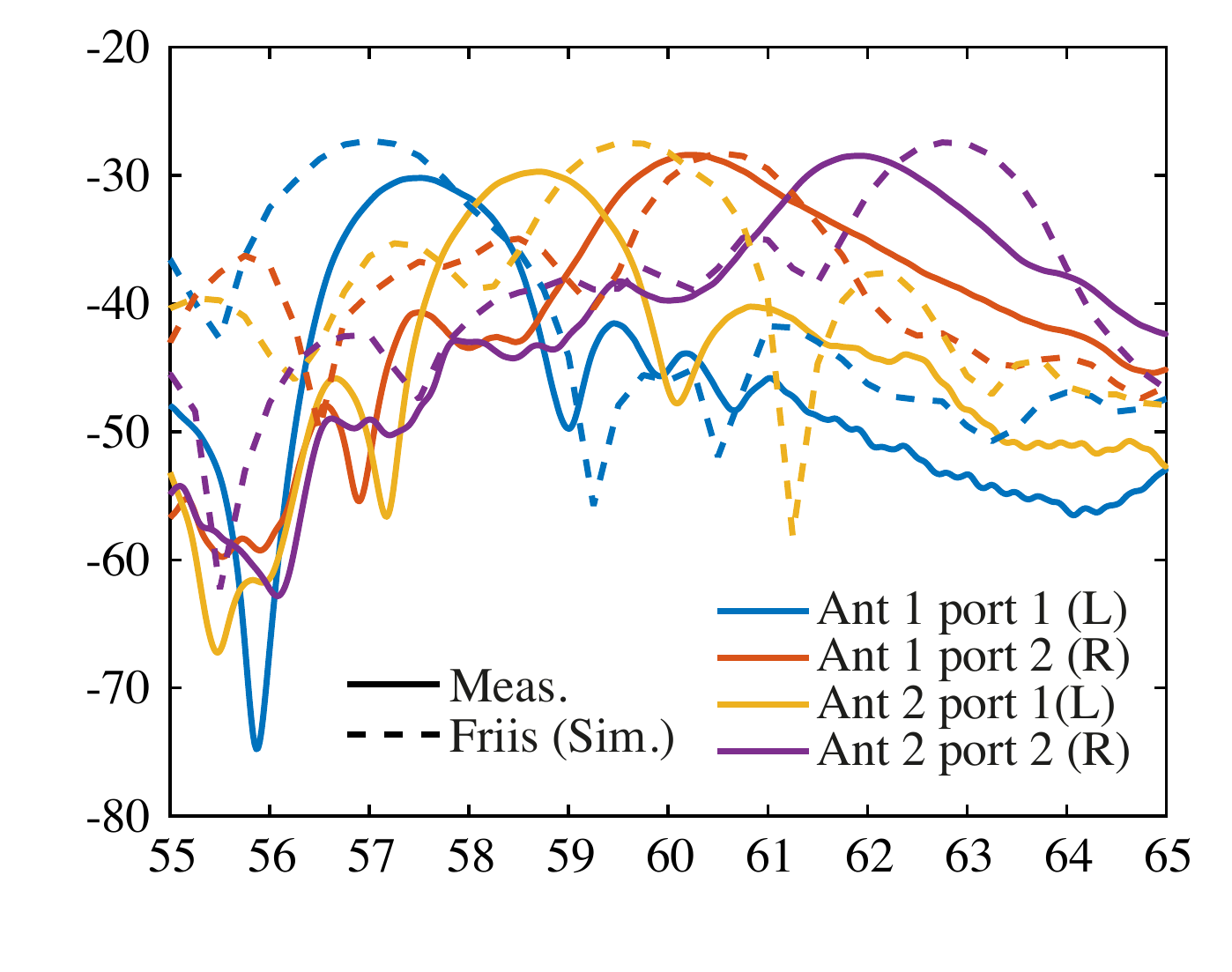}
				\put(51,4.5){\makebox(0,0){\scriptsize Frequency (GHz)}}
				\put(4.5,42){\makebox(0,0){\scriptsize\color{black}\rotatebox{90}{Transmission~(dB)}}}
		\end{overpic}
		\caption{}
	\end{subfigure}\\%
	\begin{subfigure}{\textwidth}
		\begin{overpic}[grid=false, scale = 0.28]{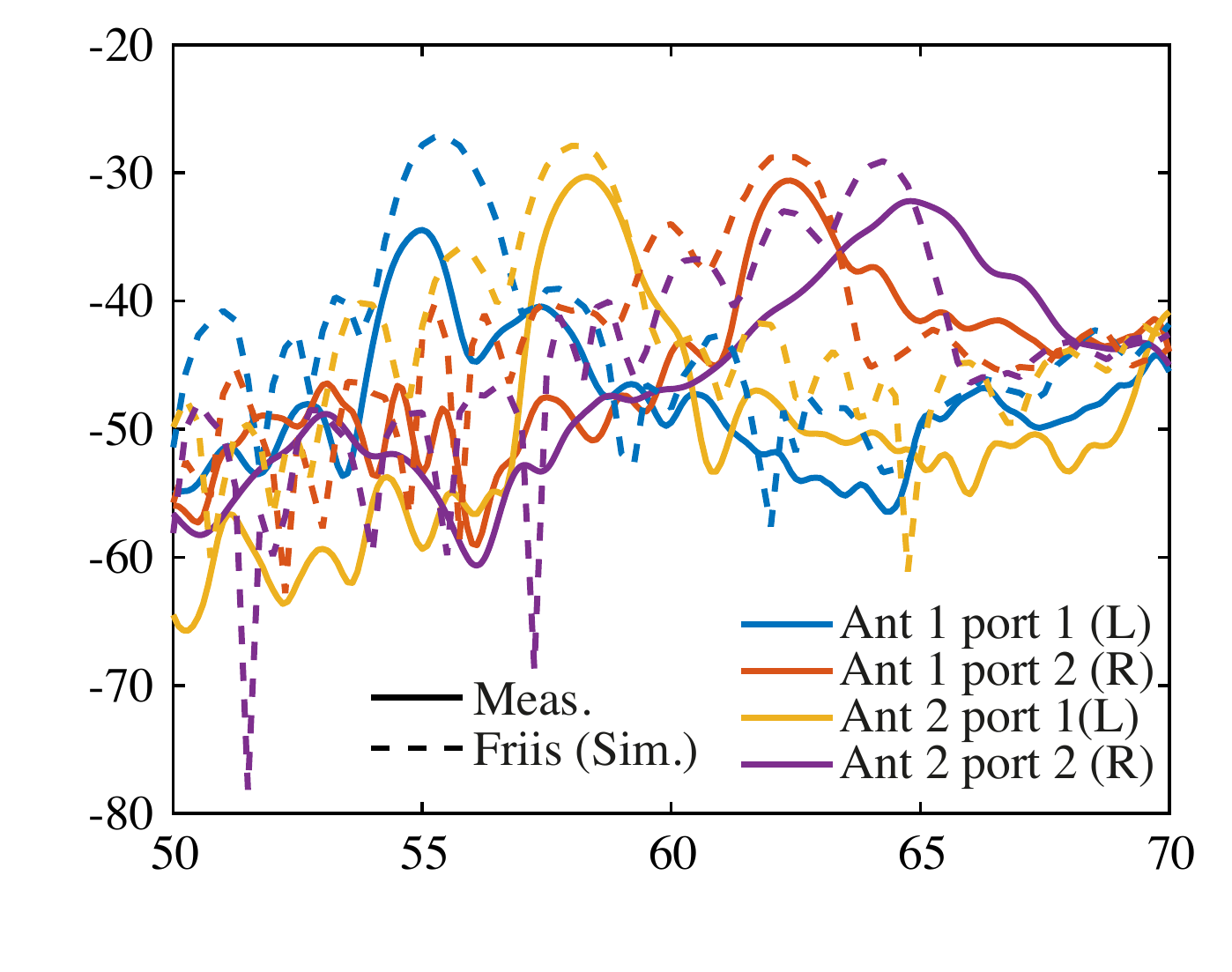}
			\put(51,4.5){\makebox(0,0){\scriptsize Frequency (GHz)}}
			\put(4.5,42){\makebox(0,0){\scriptsize\color{black}\rotatebox{90}{Transmission~(dB)}}}
		\end{overpic}
		\caption{}
	\end{subfigure}%
	\end{minipage}
	\caption{Experimental demonstration of a Horn-LWA system as a 2- and 4-channel demultiplexer. a) Experimental setup of horn-to-LWA test. Experimental results for horn-to-LWA test of LWA 1 with the horn at b) $\theta \approx 5\degree$, and c) $\theta \approx 10\degree$. The separation between the antennas is 65~cm.}
	\label{fig:exp_demo2}
\end{figure}

	Next, horn-to-LWA transmission measurements were conducted using a vector network analyzer (VNA) as shown in Fig.~\ref{fig:exp_demo2}(a). This is used to demonstrate LWA demultiplexing of an incident plane wave. The horn was used as a transmitter at an angle $\theta$ with respect to broadside direction and LWA \#1 and LWA \#2 as receivers, i.e. 4-port antenna receiver. The horn is mounted on a custom build setup made of styrofoam forming an approximate arc, and various angles were recorded using visually guided measurements, for simplicity. A frequency sweep is then used to produce a transmitting signal at the horn and the received signals at the 4 ports of the LWAs were measured.

	Two experiments were next conducted with the horn at $\theta \approx 5\degree$ and $\theta \approx 10\degree$ and the results are shown in Fig.~\ref{fig:exp_demo2}(b) \& (c), respectively. Friis equations were further used to calculate the received power using the \textit{simulated} gain patterns\footnote{The measured gain patterns were only available until 62 GHz, which were not sufficient for broadband demonstration of the demultiplexing operation. Hence simulated patterns were used.} and the computed results are compared with those from experiments as shown in Fig.~\ref{fig:exp_demo2}(b) \& (c). As expected, the $N=2$ LWAs with $2N = 4$ ports demultiplex the wide-band signal into $2N = 4$ channels. While the theoretical Friis based calculations show a moderate agreement with the experiments, there are significant frequency shifts and differences. Several factors are responsible for this. Firstly, the simulated and measured radiation patterns have appreciable differences due to fabrication tolerances and possible errors in measurements, as explained earlier. A dominant cause of error is the rudimentary experimental setup that is used, where the supporting styrofoam structure is sensitive to cable movements causing significant angular misalignments. Moreover, the angle measurements were also prone to observation error. A more rigid experimental setup with precise angular measurement mechanism will significantly reduce these errors.

\subsection{LWA-to-LWA Multiplexing/Demultiplexing}
	
\begin{figure}[htbp]		
		\begin{minipage}{0.45\columnwidth}
		\begin{subfigure}{\textwidth}
			\centering
			\begin{overpic}[grid=false, scale = 0.06]{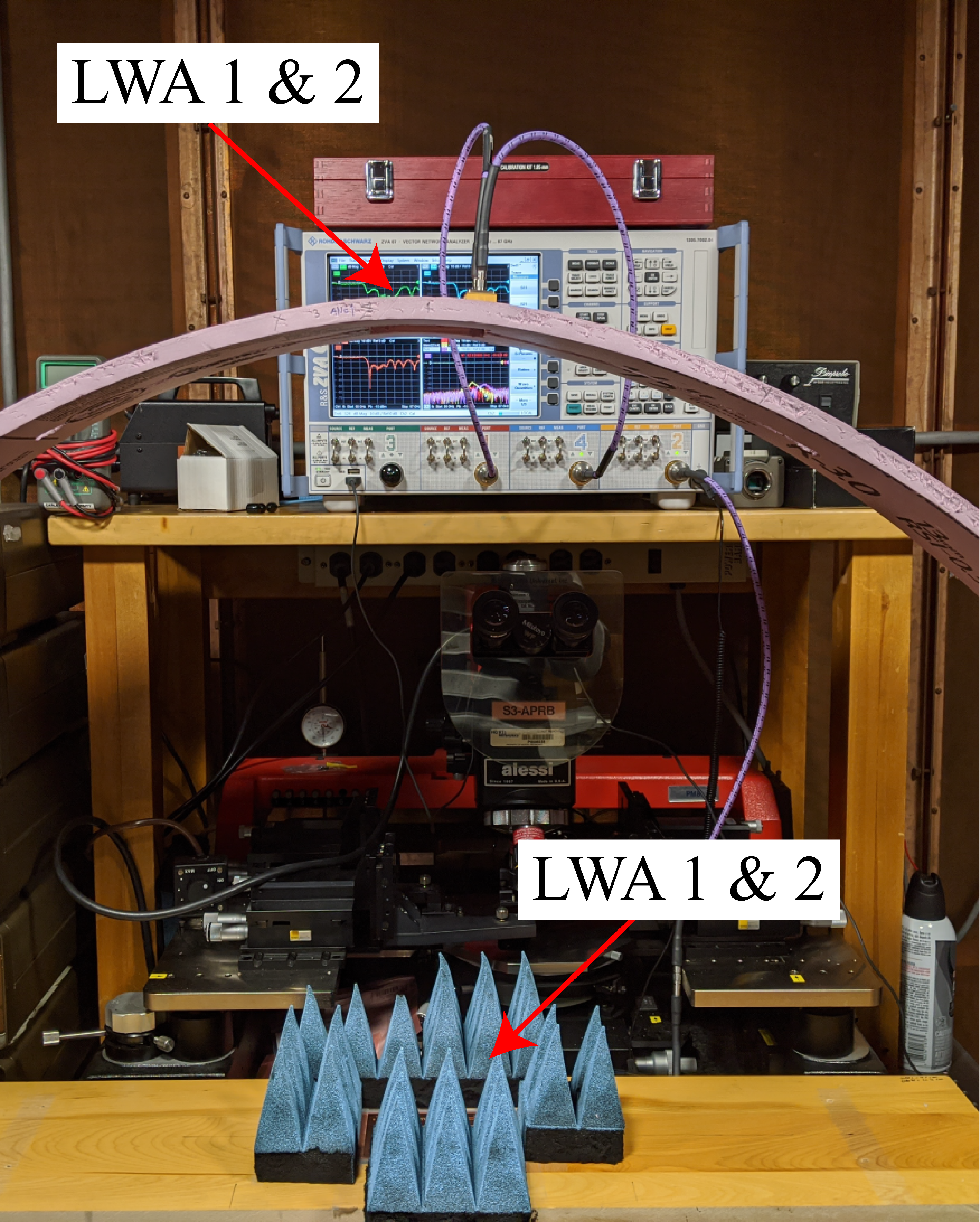}
			\end{overpic}
			\caption{}
		\end{subfigure}%
		\end{minipage}\hspace{1.2cm}
	      \begin{minipage}{0.45\columnwidth}
	      \begin{subfigure}{\textwidth}
			\begin{overpic}[grid=false, scale = 0.28]{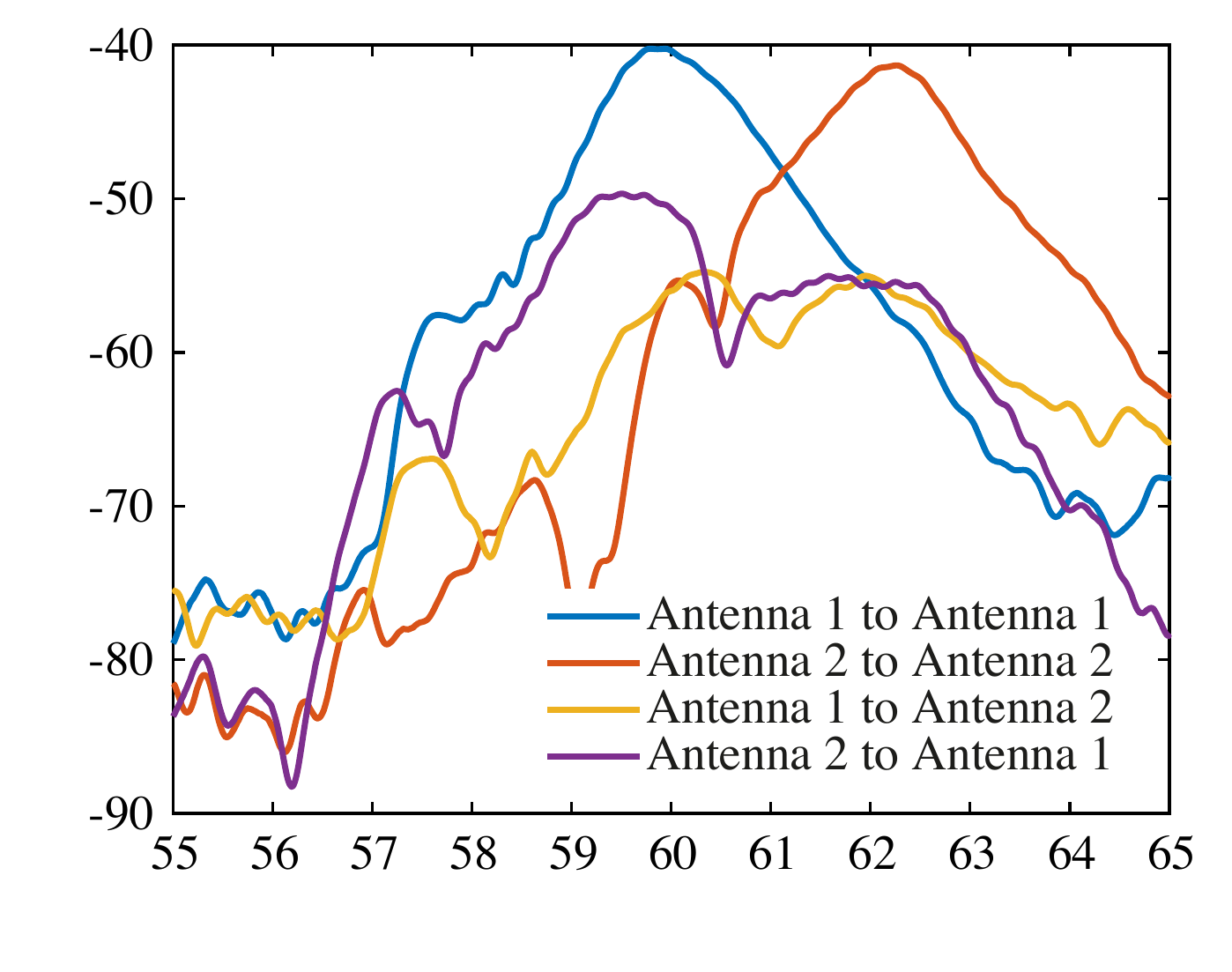}
				\put(51,4.5){\makebox(0,0){\scriptsize Frequency (GHz)}}
				\put(4.5,42){\makebox(0,0){\scriptsize\color{black}\rotatebox{90}{Transmission~(dB)}}}
			\end{overpic}
			\caption{}
		\end{subfigure} \\%
		\begin{subfigure}{\textwidth}
			\begin{overpic}[grid=false, scale = 0.28]{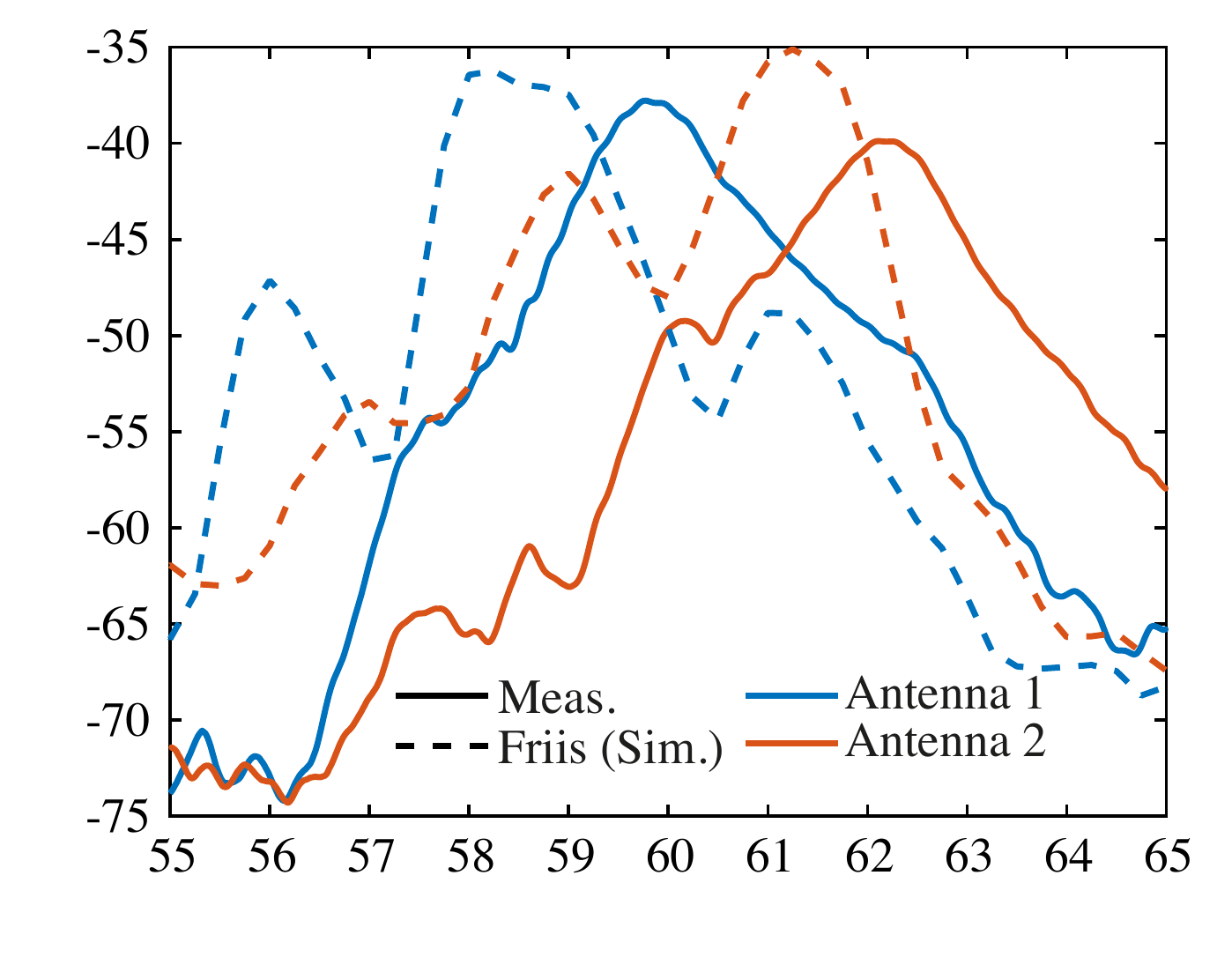}
				\put(51,4.5){\makebox(0,0){\scriptsize Frequency (GHz)}}
				\put(4.5,42){\makebox(0,0){\scriptsize\color{black}\rotatebox{90}{Transmission~(dB)}}}
			\end{overpic}
			\caption{}
		\end{subfigure}%
\end{minipage}			
		\caption{Experimental demonstration of a LWA-LWA system as a 2-channel multiplexer/demultiplexer. a) Experimental setup of LWA-to-LWA test. b) S-parameter experimental results for LWA-to-LWA test of four LWAs, two as transmitters (LWA 1 and 2) and two as receivers (LWA 1 and 2). c) Results of post-summation of S-parameters at each antenna. The separation between the antennas is 65~cm.}
		\label{fig:exp_demo3}
	\end{figure}

	Next, LWA-to-LWA transmission measurements were conducted with the antennas radiating and receiving along broadside direction at $\theta = 0\degree$, as shown in Fig.~\ref{fig:exp_demo3}(a). The VNA can only send a constant power profile therefore the frequency selective property analytically calculated in Fig.~\ref{fig:LWA_LWA_Friis}(a) will be experimentally demonstrated. Two identical pairs of the LWAs were used in this experiment. One pair (top) was used as transmitter (Tx) and the other pair (bottom) was used as receiver (Rx), i.e. 2-ports as Tx, and 2-ports as Rx. An illustration of this configuration is shown in Fig.~\ref{fig:LWA_LWA_Friis}(b). Three VNA ports were used in this experiment: two ports (P1 and P2) were permanently connected to the Tx LWAs' ports while the third port (P3) was connected to either port of the Rx LWAs to measure the received \textit{relative} power. Transmission S-parameters were next measured between all ports: $S_{31}$ \& $S_{32}$ with P3 connected to Rx LWA 1, and $S_{31}$ \& $S_{32}$ with P3 connected to Rx LWA 2. The results are shown in Fig.~\ref{fig:exp_demo3}(b). 
	
	The S-parameters for each Rx port were then summed as a post-processing step, to calculate the total relative power received at each antenna and the results are shown in Fig.~\ref{fig:exp_demo3}(c). As expected, the peak frequencies received ($59.8$~GHz \& $62.1$~GHz) were found to be close to the broadside frequencies ($59$~GHz \& ~$61$~GHz) shown in the radiation patterns \ref{fig:exp_demo1}(c) \& (d). Friis equation was further used to calculate the theoretical power received based on the simulated gain patterns and the results are also shown in \ref{fig:exp_demo1}(c). Again, only a moderate agreement is seen between measurements and theoretical calculations, while the qualitative frequency multiplexing and demultiplexing using the Tx and Rx pairs is correctly obtained. As before, the discrepancy between measurements and Friis transmission equation based computation is partly related to the dielectric constant and loss tangent variations. In addition, it is due to non-idealities in the setup with misalignment of the antennas, and the PCB warping due to the VNA cables causing stress on the WR-15 feeds. Additionally, since the radiation pattern was measured for only one pair, while Friis transmission assumes the two LWA pairs are identical, which is likely not the case due to the fabrication tolerances between the two PCBs.

\section{LWA Bandwidth Investigation \& Discussion}

The heart of the proposed system is the LWA antenna used in achieving the demultiplexing and multiplexing operation. The exact channel bandwidth and the placement of the channel frequencies are governed by the properties of the LWA used. Precisely, the channel bandwidth supported by the antenna depends on the beamwidths of the antenna, and can be controlled either by changing the leakage factor $\alpha(\omega)$ or/and by controlling the physical length of the antenna. The leakage-factor requires a redesign of the LWA period (e.g. the reflection cancelling slot pair with different slot lengths, separations and offset), while the physical length of the antenna can be increased by simply cascading more number of unit cells.
	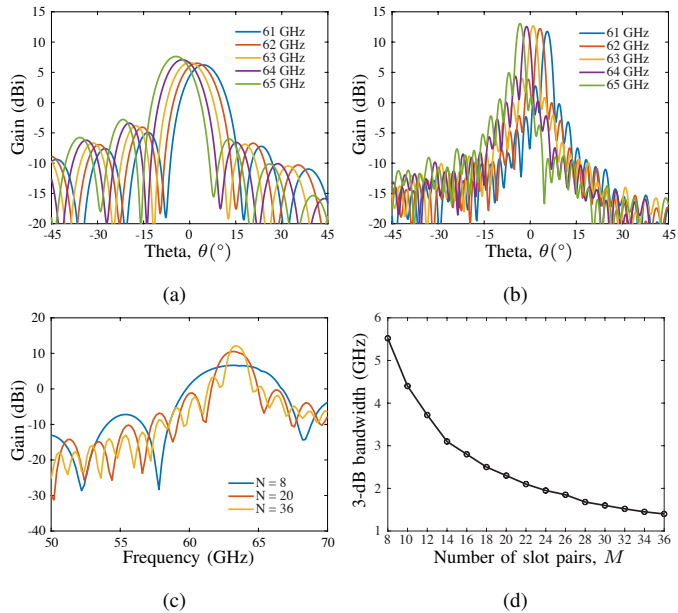
\begin{figure}[htbp]
		\centering
		\begin{subfigure}{0.25\textwidth}
			\centering
			\begin{overpic}[grid=false, scale = 0.32]{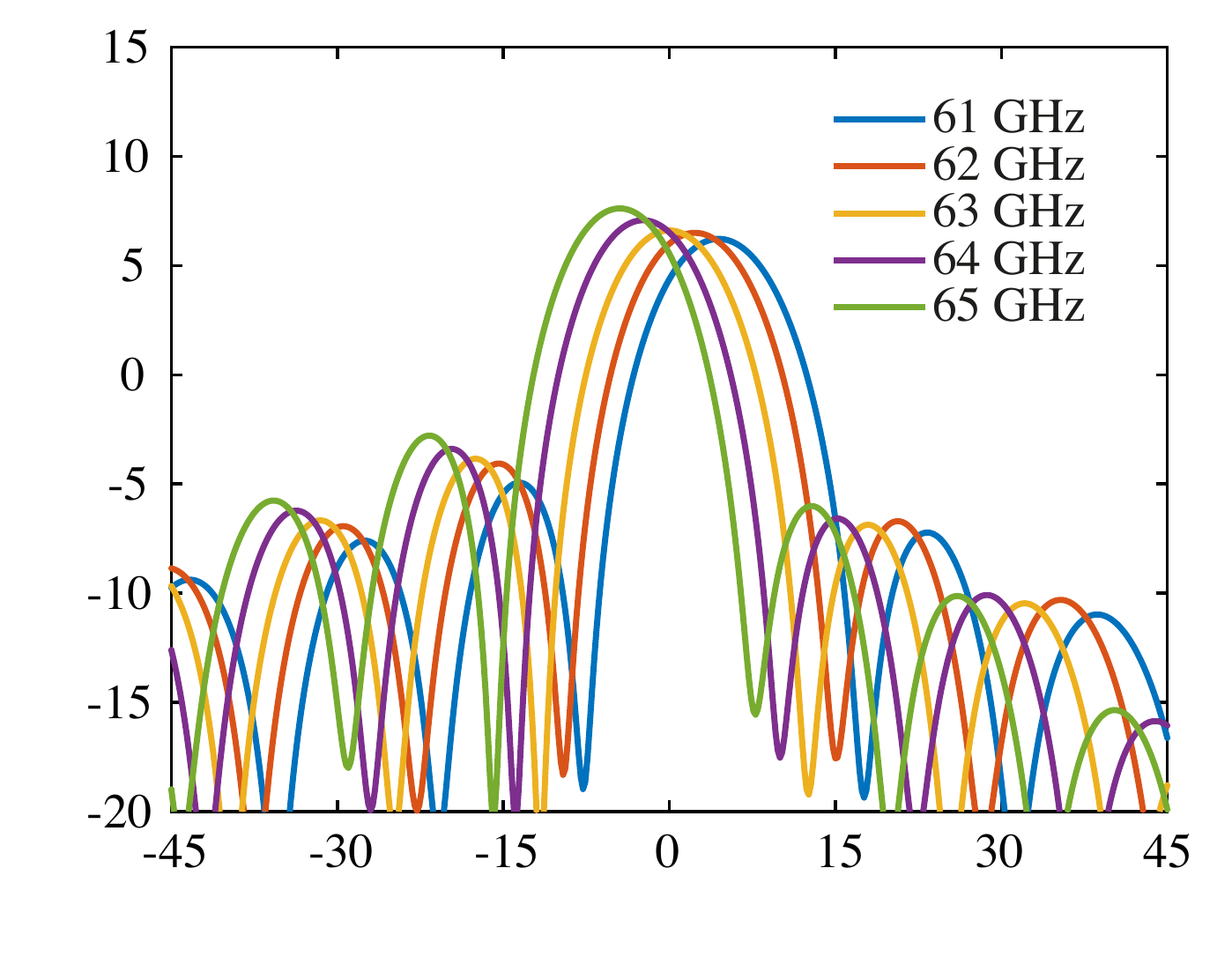}
				\put(54,4.5){\makebox(0,0){\scriptsize Theta, $\theta(\degree)$}}
				\put(4.5,45){\makebox(0,0){\scriptsize\color{black}\rotatebox{90}{Gain~(dBi)}}}
			\end{overpic}
			\caption{}
		\end{subfigure}%
		\begin{subfigure}{0.25\textwidth}
			\centering
			\begin{overpic}[grid=false, scale = 0.32]{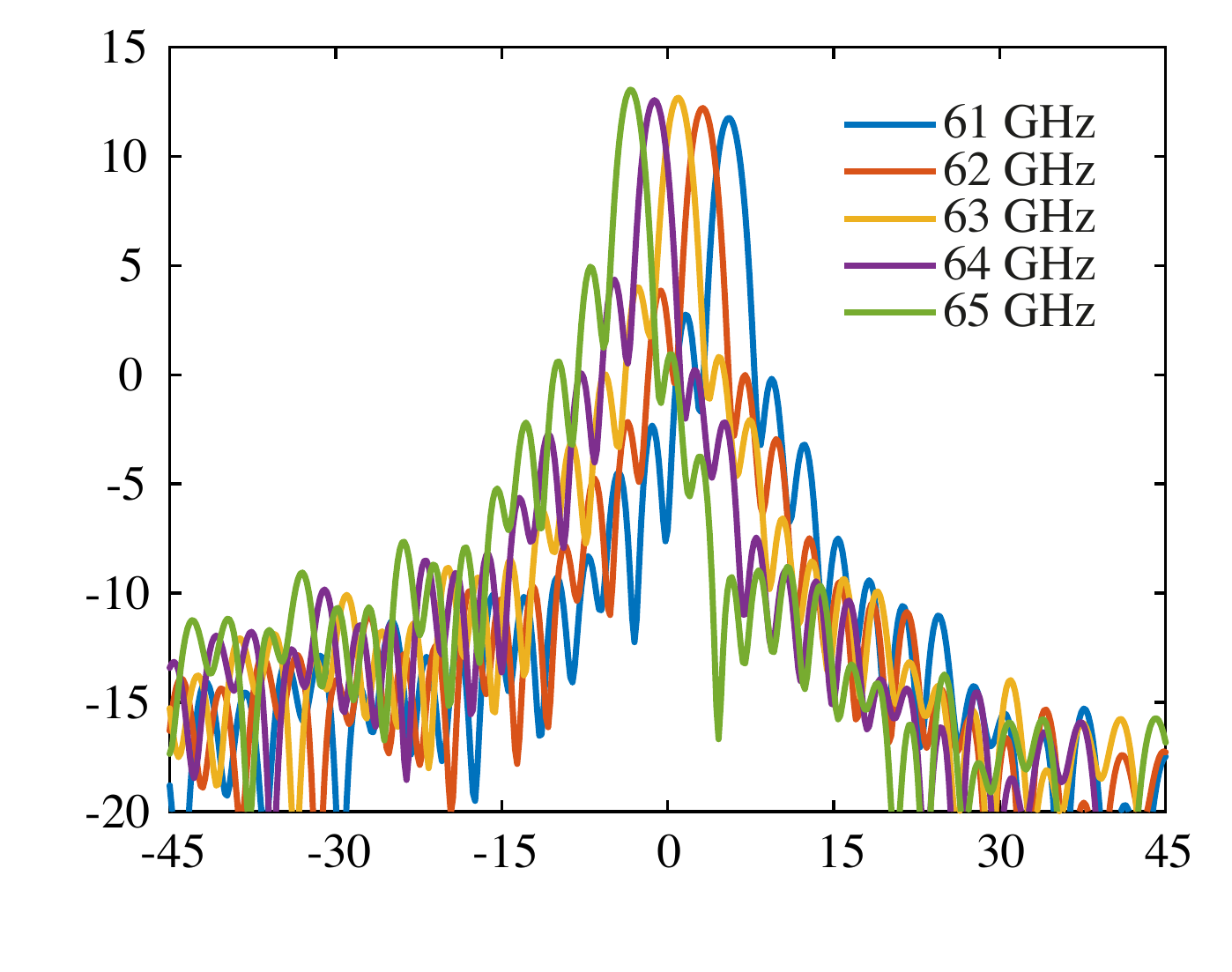}
				\put(54,4.5){\makebox(0,0){\scriptsize Theta, $\theta(\degree)$}}
				\put(4.5,45){\makebox(0,0){\scriptsize\color{black}\rotatebox{90}{Gain~(dBi)}}}
			\end{overpic}
			\caption{}
		\end{subfigure}%
	
		\begin{subfigure}{0.25\textwidth}
		\centering
		\begin{overpic}[grid=false, scale = 0.32]{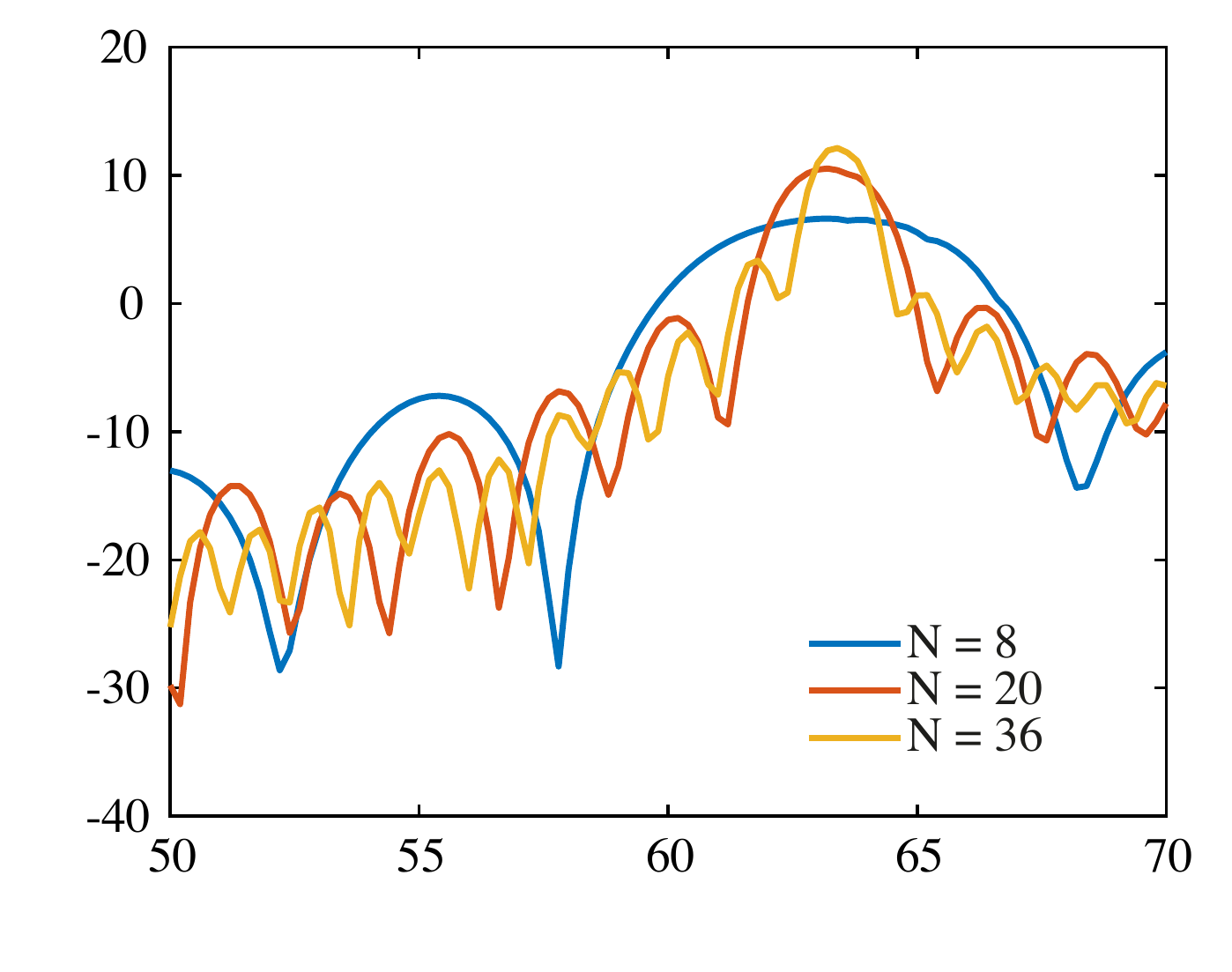}
			\put(54,4.5){\makebox(0,0){\scriptsize Frequency~(GHz)}}
			\put(4.5,45){\makebox(0,0){\scriptsize\color{black}\rotatebox{90}{Gain~(dBi)}}}
		\end{overpic}
		\caption{}
		\end{subfigure}%
		\begin{subfigure}{0.25\textwidth}
			\centering
			\begin{overpic}[grid=false, scale = 0.32]{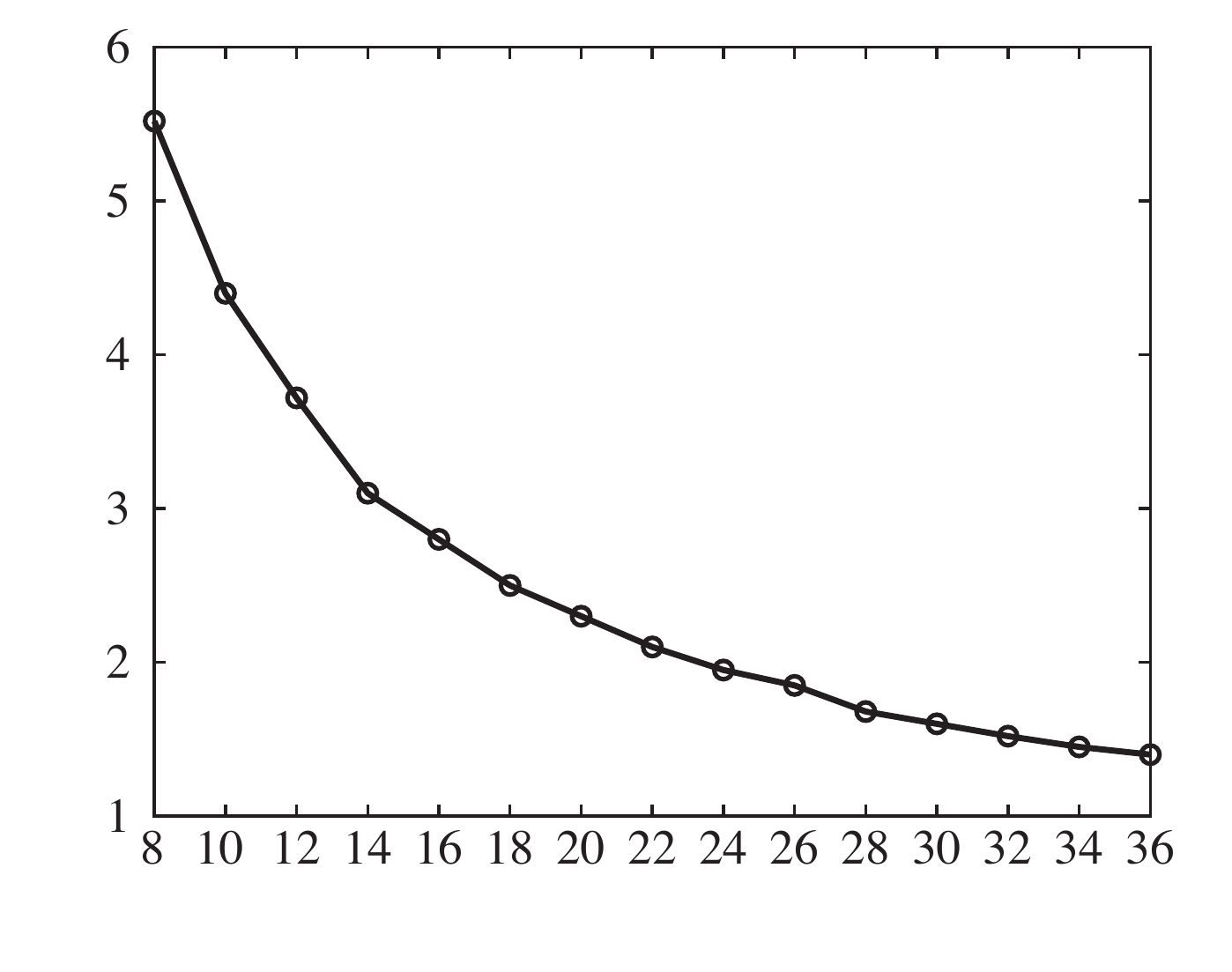}
				\put(54,4.5){\makebox(0,0){\scriptsize Number of slot pairs, $M$}}
				\put(4.5,45){\makebox(0,0){\scriptsize\color{black}\rotatebox{90}{3-dB bandwidth~(GHz)}}}
			\end{overpic}
			\caption{}
		\end{subfigure}%
	
		\caption{Investigation of the effect of antenna length on the channel bandwidth supported by a LWA with period, $p=2.78$~mm. a) Radiation patterns with $M=8$ slots, b) radiations patterns with $M=36$ slots, c) gain versus frequency at broadside for $M=8$, $20$, and $36$, and d) computed $3$-dB bandwidth versus number of slot pairs. In each case, the length of the antenna is $l=M\times p$.}
		\label{fig:LWA_bandwidth}
	\end{figure}

To illustrate this property, Fig.~\ref{fig:LWA_bandwidth}(a) \& (b) show the far-field radiation patterns of two antenna lengths, exhibiting different beamwidths. The longer antenna has a narrower beamwidth and vice versa. Fig.~\ref{fig:LWA_bandwidth}(c) further shows the gain versus frequency at broadside for three antenna lengths, showing the variation in the bandwidth. The 3-dB bandwidth of these profiles centered at the peak may be considered as the channels supported by the antenna. Consider these antennas receiving power from a broadband source, and thus acting as demultiplexers. A short antenna ($M=8$) accepts a large frequency bandwidth because its beamwidth is large and thus has a large acceptance angular range. As the antenna becomes longer, and the pattern becomes directive, the frequency bandwidth accepted by the antenna becomes narrower, as clearly seen with LWA of $M=20$ and $M=36$ cells. Fig.~\ref{fig:LWA_bandwidth}(d) further shows the $3$-dB bandwidth as a function of the number of slot pairs $M$. The channel bandwidth monotonically decreases with increasing number of slot periods as it gradually becomes more stable for large lengths. In this example, there is little change in the bandwidth beyond $M=30$, for instance which is expected since, as the antenna gets longer, more power is radiated until very little power is left at the end of aperture. Any further increase in the length subsequently has little impact. 

The exact placement of the channel frequencies is governed by the beam-scanning law of the antenna and the slot period $p$ which can be used to precisely point a channel to a desired directions. While uniform slots were used for the demonstration here, more control over the aperture may be achieved by designing non-uniform slots and using integrated waveguides of varying widths to control radiation. This can further allow us to control the separation between the bands for each antenna. In addition, while a linearly-polarized antenna was proposed here, a circularly-polarized SIW slot array antenna can be designed using a similar architecture for adding polarization diversity \cite{Chiang_Circular_Slot, Min_U_Slot, Kai_Circular_Slot, Cheng_Quadri_Polarization, Jing_Circular_Slot, Hirokawa_Circular_Slot}. 

\section{Conclusion}
	
	A novel multiplexing/demultiplexing technique using an array of LWAs forming a multi-port antenna aperture has been proposed and experimentally demonstrated in the mm-wave band at $60$~GHz. The system is based on the frequency discriminating capability of LWAs in the form of the frequency-dependent beam-scanning law. It has been shown that various ports of the multi-port antenna structure, when excited with an incoming broadband plane-wave along a specific angle, are mapped to specific channel frequencies, so that the LWA array act as a channel demultiplexer.	 Following reciprocity, when the same ports are excited with different but strategically chosen channel frequencies, the antenna aperture radiates all channels along a single direction in space, effectively multiplexing the channels. Furthermore, this concept has been extended to a LWA-to-LWA Tx-Rx system, where $2N$ frequency channels are multiplexed and demultiplexed  using $N$ antenna pairs communicating in LOS conditions. The principle of the system has been demonstrated using a simple analytical model based on Friis transmission equation followed by a proof-of-concept experiment showing a horn-to-LWA and LWA-to-LWA based channel discrimination using reflection cancelling slot antenna arrays. The proposed solution thus rests on engineered LWA arrays which are low-profile, compact, easy to fabricate and do not require any matching networks, as typically required in conventional multiplexing/demultiplexing techniques. Furthermore, compatible with SIW technology, the proposed solution is easy to integrate with other front-end circuitry such as low-noise amplifiers in a typical communication system chain. The proposed design thus opens the door for interesting avenues to be explored for designing high performance mm-wave antenna front-ends for wireless communication in the 5G and future wireless applications.

\bibliographystyle{ieeetran}
\bibliography{2020_TAP_LWA_MUX_DEMUX}

\end{document}